\documentclass[aps,prd,showpacs,preprintnumbers,superscriptaddress,twocolumn]{revtex4-2}
\usepackage{amsmath,amssymb,amsfonts}
\usepackage{graphicx}
\usepackage{hyperref}
\usepackage{float}

\begin{document}
	
\title{Momentum Exchange in the Dark Sector: Cosmological Consequences and Radiative Neutrino Masses}
	
\author{Mohid Farhan}
\affiliation{Department of Space Science, Institute of Space Technology, Islamabad, Pakistan}
	
\date{\today}
	
\begin{abstract}
We extend a symmetry-protected dark sector to include active neutrino interactions within a $Z_4 \times Z_2^{\rm dark}$-symmetric inert doublet plus singlet framework. Dark matter is stabilized by the exact $Z_2^{\rm dark}$ symmetry, while dark energy arises as a radiatively stable pseudo-Nambu-Goldstone boson from a softly broken global $U(1)_S$. Neutrino masses are generated radiatively through the scotogenic seesaw mechanism. We introduce two technically natural derivative couplings: one between the dark energy scalar and the neutrino energy current, and one between the scalar dark matter field and the neutrino energy current. Together with the DM--DE operator, this creates a unified dark sector with three momentum-exchange channels. We derive the covariant fluid equations and implement the interactions in \texttt{CLASS}. Each channel leaves a distinct cosmological imprint. The DM--DE drag suppresses $\sigma_8$ monotonically, saturating at $-2\%$. The DM--$\nu$ channel is non-monotonic: it enhances $\sigma_8$ by up to $7.4\%$ at $g_{\chi\nu} \simeq 3\times 10^{-4}$, before suppressing it at larger coupling. While the suppressing regime could alleviate the $S_8$ tension, the entire parameter space above \(g_{\chi\nu} \lesssim 10^{-5}\) is excluded by the small-scale matter power spectrum. However, the DM--$\nu$ channel leaves a distinctive enhancement of the CMB lensing potential, reaching $\sim 5\%$ at $\ell \sim 1000$ for viable couplings and growing to $\sim 17\%$ at $\ell \sim 500$ for the $\sigma_8$-peak coupling; this signal is within reach of CMB-S4. We demonstrate that the DM--$\nu$ coupling and the total neutrino mass modify $\sigma_8$ independently, breaking their degeneracy. Ultimately, the \((1+z)^6\) scaling of the derivative coupling forces tight coupling at high redshifts, ruling out the \(S_8\)-resolving regime.
\end{abstract}
\maketitle

\section{Introduction}
\label{sec:intro}

The nature of dark matter (DM) and dark energy (DE) remains one of the most profound open problems in modern cosmology, dating back to early dynamical evidence of unseen mass \cite{Zwicky1933} and the late 1990s Type Ia supernovae observations establishing cosmic acceleration \cite{Riess1998, Perlmutter1999}. A wealth of astrophysical and cosmological observations—from galaxy rotation curves to the cosmic microwave background (CMB)—establish that DM constitutes approximately $27\%$ of the cosmic energy budget and plays a crucial role in structure formation~\cite{Bertone2005}. Similarly, the observed late-time accelerated expansion of the Universe, first inferred from Type Ia supernovae and subsequently confirmed by CMB and large-scale structure (LSS) data, points to a DE component comprising roughly $68\%$ of the energy density~\cite{Copeland2006}. The microscopic nature of neither component is understood within the Standard Model of particle physics \cite{Workman2022}.

A theoretically compelling possibility is that DM and DE are not entirely decoupled but interact with each other \cite{Wetterich1995, Amendola2000, Farhan2025}. Interacting dark energy (IDE) models in which DM and DE exchange momentum without background energy transfer are particularly attractive because they leave the primary CMB almost unchanged~\cite{Simpson2010, Farhan:2026sym}. Such models are technically realized by Type-3 fluid--scalar couplings $F(A)$, $A=u_X^\mu\partial_\mu\phi$, where the scalar field's gradient is contracted with the DM four-velocity~\cite{Pourtsidou2013}. Crucially, the Type-3 class is distinguished by the fact that the interaction is a pure momentum exchange: the background energy densities evolve as in the uncoupled case, and the coupling is radiatively stable because it is protected by the shift symmetry of the scalar field~\cite{Farhan2025,Pourtsidou2013,Farhan:2026sym}. This radiative naturalness is a decisive advantage over energy-exchange couplings, which generically require non-derivative operators that violate the shift symmetry and reintroduce radiative instabilities \cite{Coleman1973, Farhan:2026no-go}, while recently constructed exceptions exist \cite{Bagherian2026}. The perturbative structure of Type-3 models is understood through the parameterized post-Friedmannian framework~\cite{Skordis2015,Koivisto2015}, and they have been implemented in the Boltzmann solver \texttt{CLASS}~\cite{Pourtsidou2016}, with later work studying variable scalar sound speed~\cite{Linton2018} and nonlinear prescriptions~\cite{Carrilho2022}.

Neutrinos, while part of the Standard Model, act as an equally elusive and fundamental component linked to dark sector phenomenology, leaving distinct imprints on cosmic structure formation through their free-streaming effects \cite{Bond1980, Lesgourgues2006}. The discovery of neutrino oscillations established that neutrinos have non-zero masses, requiring physics beyond the Standard Model~\cite{King2015}. The absolute neutrino mass scale and the ordering of the mass eigenstates—the neutrino mass hierarchy—remain unknown, representing one of the major gaps in our understanding of particle physics~\cite{King2015}. A particularly elegant solution to the hierarchy problem is the Dirac mass cancellation mechanism, in which the large Dirac mass term expected from grand unified theories is partially or completely cancelled in the light neutrino mass matrix by a discrete flavor symmetry~\cite{DiracCancellation2009}. This mechanism, and its radiative variants, provides a natural explanation for why neutrino masses are so much smaller than the charged fermion masses \cite{Cao2011}. In the context of dark sectors, discrete symmetries such as \(Z_4\) and \(Z_2\) can simultaneously stabilize the dark matter candidate and forbid tree-level Dirac masses, generating neutrino masses radiatively through dark-sector loops~\cite{Bhattacharya2018,Chulia2017}.

Given the elusive nature of DM, DE, and neutrinos, it is natural to ask whether they interact with each other. Neutrino--DM interactions have been extensively studied in the literature, primarily through elastic scattering processes that induce diffusion-damped dark acoustic oscillations in the matter power spectrum~\cite{Lattanzi2014}. Neutrino--DE interactions have also been explored, particularly in the context of mass-varying neutrino models where the DE scalar field drives a time-dependent neutrino mass~\cite{Euclid2018}. These scenarios demonstrate that even weak interactions between the dark components can leave distinctive imprints on cosmological observables.

Despite the success of the $\Lambda$CDM model, persistent tensions have emerged between early-universe and late-universe measurements. The most prominent is the $S_8$ tension: the clustering amplitude inferred from Planck CMB data~\cite{Planck2018} is consistently higher than the value derived from low-redshift weak-lensing surveys such as KiDS-1000~\cite{KiDS2021} and DES Year~3~\cite{DESY3}. In parallel, local measurements of the Hubble constant $H_0$ remain in tension with CMB predictions~\cite{Riess2022}. These discrepancies motivate extensions of the dark sector \cite{Naeem:2026layered}. Interestingly, recent work has shown that phenomenological DM--neutrino elastic scattering can alleviate the $S_8$ tension by tuning the scattering cross-section to $u_{\nu\rm DM} \sim 10^{-4}$~\cite{Zu2026}. In that framework, a constant cross-section gives a drag rate scaling as $(1+z)^3$, allowing the interaction to suppress late-time growth without catastrophically affecting the early universe.

In this paper we take a different approach. We extend the $Z_4 \times Z_2^{\rm dark}$-symmetric inert doublet plus complex singlet framework~\cite{Naeem:2026layered,Farhan:2026sym} to include two technically natural neutrino interactions: a derivative coupling between the DE pNGB and the neutrino energy current, and a derivative coupling between the scalar DM field and the neutrino energy current. The couplings are suppressed by the cutoff scale $\Lambda^2$ and preserve all symmetries, including the shift symmetry of the DE field. This creates a unified dark sector with three momentum-exchange channels. We implement the model in \texttt{CLASS} and study the resulting background evolution, linear matter power spectra, primary CMB spectra, CMB lensing potential, and the growth rate $f\sigma_8(z)$. The three channels act on different pairs of species and therefore leave distinguishable signatures. The DM--DE channel behaves as in the original $Z_4$-IDSM model, producing a monotonic $\sigma_8$ suppression that saturates at the percent level. The DM--$\nu$ channel is non-monotonic, enhancing $\sigma_8$ by up to $7.4\%$ at moderate coupling and strongly suppressing it at larger coupling. The suppressing regime required to alleviate the $S_8$ tension is ruled out by the small-scale matter power spectrum, which constrains the coupling to $g_{\chi\nu} \lesssim 10^{-5}$; at viable couplings the effect on $\sigma_8$ is therefore negligible. This contrasts with the phenomenological scenarios ~\cite{Zu2026}: our derivative EFT coupling scales as $(\rho_\nu + P_\nu)H \propto (1+z)^6$ in the radiation era, three additional powers of $(1+z)$ compared to a constant cross-section, which is precisely what makes the $S_8$-resolving regime inaccessible. The $\nu$--DE channel is null in the primary CMB, $\sigma_8$, and the lensing potential even at saturation, because it does not couple to cold dark matter. The DM--$\nu$ channel, in contrast, leaves a distinctive enhancement of the CMB lensing potential, reaching $\sim 5\%$ at $\ell \sim 1000$ for viable couplings, providing a channel-specific target for future CMB experiments. This complementarity is the central observational signature of the unified framework, and it also clarifies why the extended dark sector does not resolve the $S_8$ tension: the couplings that would produce the necessary suppression are ruled out by the same small-scale data that motivates the tension in the first place.

A central feature of our covariant derivation from the perfect-fluid stress tensor is that the momentum-exchange drag rates are weighted by the fluid's enthalpy density $\rho_i + P_i$, not merely by $\rho_i$. For neutrinos, this implies a factor of $4/3$ enhancement in the early universe when they are relativistic, a robust consequence of the covariant drag structure that distinguishes our treatment from phenomenological parameterizations in which the rate is taken to be proportional to $\rho_\nu$ or treated as an arbitrary function of scale factor. For dark energy, $\rho_{\mathrm{DE}} + P_{\mathrm{DE}} \approx 0$ during slow-roll, which affects the back-reaction terms in the dark energy Euler equation.

The paper is organized as follows. In Sec.~\ref{sec:model} we define the full Lagrangian and symmetries. Section~\ref{sec:background_perturbations} gives the background evolution and perturbation equations. Section~\ref{sec:stability} analyzes stability and the tight-coupling approximation. Section~\ref{sec:method} describes the \texttt{CLASS} implementation, and Sec.~\ref{sec:results} presents the linear cosmological signatures of the three channels. Section~\ref{sec:conclusions} summarizes the results and outlines the prospects for a full MCMC analysis against current data.
	
\section{Model and Symmetries}
\label{sec:model}
	
We extend the $Z_4 \times Z_2^{\rm dark}$-symmetric inert doublet plus singlet framework ~\cite{Naeem:2026layered,Farhan:2026sym} to include active neutrino interactions. The $Z_2^{\rm dark}$ factor is introduced in this work to enforce absolute stability of the dark matter candidate; it acts only on the inert doublet and the right-handed neutrinos, while leaving the singlet sector invariant. It therefore acts independently of (and trivially commutes with) the global $U(1)_S$ and \(Z_4\) symmetries that govern the dark energy and seesaw sectors. The field content consists of:
	
	\begin{itemize}
		\item the Standard Model Higgs doublet $H_1$,
		\item an inert scalar doublet $H_2$ with dark matter candidate $H_2^0\equiv\chi$,
		\item a complex scalar singlet $S$,
		\item the Standard Model lepton doublets and right-handed charged leptons,
		\item three right-handed neutrino singlets $N_{R_i}$,
		\item together with active massive neutrinos.
	\end{itemize}
	
    The theory is endowed with a discrete $Z_4 \times Z_2^{\rm dark}$ symmetry and a global $U(1)_S$ symmetry. Under $Z_4$, the fields transform as
    \begin{equation}
	H_1 \to H_1,\qquad
	H_2 \to -H_2,\qquad
	S \to iS,\qquad
	N_{R_i} \to -N_{R_i},
	\label{eq:z4_transformations}
    \end{equation}
    while under $Z_2^{\rm dark}$ the transformation rules are
    \begin{equation}
	H_1 \to H_1,\qquad
	H_2 \to -H_2,\qquad
	S \to S,\qquad
	N_{R_i} \to -N_{R_i}.
	\label{eq:z2dark_transformations}
    \end{equation}
    Under the global $U(1)_S$, the singlet field transforms as
    \begin{equation}
	S \to e^{i\alpha} S,
	\label{eq:u1_transformation}
    \end{equation}
    while the SM Higgs doublet $H_1$, the inert doublet $H_2$, and all Standard Model fields carry zero $U(1)_S$ charge and transform trivially. These symmetries are independent: $Z_4$ acts on the dark sector fields to enable the scotogenic seesaw mechanism, $Z_2^{\rm dark}$ acts on $H_2$ and $N_{R_i}$ to enforce absolute stability of the dark matter candidate, and $U(1)_S$ acts only on the singlet $S$ to protect the pNGB dark energy mass. We note that the introduction of an independent \(Z_2^{\rm dark}\) symmetry is a necessary extension of the original \(Z_4\) framework~\cite{Naeem:2026layered,Farhan:2026sym}. In the original model, the soft-breaking term \(\mu_{sb}^2 S^2\) explicitly broke \(Z_4\) down to a residual \(Z_2\), under which \(H_2\) was even. While the decay of \(H_2\) into Standard Model states was heavily suppressed by the tiny soft-breaking scale, rendering it effectively stable for cosmological purposes, the addition of the scotogenic Yukawa coupling \(y_{ij} \bar{L}_L \tilde{H}_2 N_R\) opens a new decay channel into leptons and neutrinos. To guarantee that \(H_2^0\) is absolutely stable and to ensure the rigor of the scotogenic mechanism, we introduce the exact \(Z_2^{\rm dark}\) symmetry, which acts independently on \(H_2\) and \(N_R\), leaving the \(S\)-sector dynamics untouched.
	
	Upon spontaneous breaking of $U(1)_S$ by $\langle S \rangle = v_s/\sqrt{2}$, the complex scalar $S$ decomposes into a heavy radial mode $\rho$ and a light pseudo-Nambu-Goldstone boson $\phi$. In the limit where the soft-breaking term $\mu_{sb}^2$ vanishes, the Goldstone mode $\phi$ would enjoy an exact shift symmetry. The nonzero but small $\mu_{sb}^2$ breaks this shift symmetry softly, generating the pNGB potential and giving $\phi$ its tiny mass $m_\phi \sim H_0$. This ensures that the dark energy mass is technically natural: in the limit $\mu_{sb}^2 \to 0$, the shift symmetry is restored and the mass vanishes. Moreover, the shift symmetry (even when softly broken) forbids non-derivative couplings of $\phi$ that would generate large radiative corrections, which is why only derivative interactions of $\phi$ are allowed.
	
\subsection{Scalar potential and pNGB protection condition}
\label{sec:potential}

The most general renormalizable scalar potential consistent with gauge invariance and the $Z_4$ charges (before imposing $Z_2^{\rm dark}$ and $U(1)_S$) is
\begin{equation}
	\begin{split}
		V_{\rm full} &=
		\mu_1^2|H_1|^2
		+\mu_2^2|H_2|^2
		+\frac{\lambda_1}{2}|H_1|^4
		+\frac{\lambda_2}{2}|H_2|^4 \\
		&\quad
		+\lambda_3|H_1|^2|H_2|^2
		+\lambda_4|H_1^\dagger H_2|^2 
		+\left[\frac{\lambda_5}{2}(H_1^\dagger H_2)^2+\text{h.c.}\right] \\
		&\quad
		+\mu_S^2|S|^2
		+\frac{\lambda_S}{2}|S|^4
		+\left[\frac{\mu_{sb}^2}{2}S^2+\text{h.c.}\right] \\
		&\quad
		+\lambda_{S1}|S|^2|H_1|^2
		+\lambda_{S2}|S|^2|H_2|^2
		+\lambda_{S12}S^2(H_1^\dagger H_2) \\
		&\quad
		+\lambda_{S21}(S^\dagger)^2(H_2^\dagger H_1)
	\end{split}
	\label{eq:full_potential}
\end{equation}

The last line contains singlet--doublet portal operators. These operators fall into two classes:

\begin{itemize}
	\item \textbf{Class I:} $|S|^2|H_1|^2$ and $|S|^2|H_2|^2$. In the polar parameterization $S = (v_s + \rho)e^{i\phi/v_s}/\sqrt{2}$, the absolute square is $|S|^2 = (v_s + \rho)^2/2$, which depends only on the radial mode $\rho$. Consequently, these operators are manifestly invariant under the shift symmetry $\phi \to \phi + c$ (as well as $U(1)_S$, $Z_4$, and $Z_2^{\rm dark}$). They neither generate a mass term for the pNGB nor destabilize the dark matter. After integrating out the heavy radial mode $\rho$, they generate derivative operators such as $|H_i|^2 \partial_\mu \phi \partial^\mu \phi$, which are also shift-symmetric. Their absence is therefore not enforced by any symmetry; it is a phenomenological restriction that isolates the pure momentum-exchange limit studied in this work. Lifting this restriction is deferred for future work. 
	
	\item \textbf{Class II:} $S^2(H_1^\dagger H_2)$ and $(S^\dagger)^2(H_2^\dagger H_1)$. These operators are neither $U(1)_S$-invariant nor $Z_2^{\rm dark}$-invariant, and they explicitly break the shift symmetry through factors of $e^{\pm 2i\phi/v_s}$. They are forbidden by the exact $Z_2^{\rm dark}$ symmetry and cannot be generated at any loop order, since $Z_2^{\rm dark}$ is exact and is not broken by the soft term $\mu_{sb}^2 S^2$. As an independent check, even if $Z_2^{\rm dark}$ were relaxed, the one-loop correction to the pNGB mass
	\begin{equation}
		\delta m_\phi^2
		\sim
		\frac{\lambda_{S12}^2}{16\pi^2}
		m_{H_2}^2
		\ln\left(\frac{\Lambda^2}{m_{H_2}^2}\right)
		\label{eq:delta_mphi}
	\end{equation}
	would require $\lambda_{S12} \lesssim 10^{-44}$ for $m_{H_2} \sim 10^2$~GeV and $m_\phi \sim H_0 \sim 10^{-42}$~GeV, so these operators are technically natural to be small in any case.
\end{itemize}

We therefore impose the pNGB protection condition
\begin{equation}
	\lambda_{S1}=\lambda_{S2}=\lambda_{S12}=\lambda_{S21}=0 ,
	\label{eq:pNGB_protection}
\end{equation}
where $\lambda_{S1}$ and $\lambda_{S2}$ are set to zero as a phenomenological choice (Class I), and $\lambda_{S12}$ and $\lambda_{S21}$ vanish by $Z_2^{\rm dark}$ (Class II).
With these couplings set to zero, the remaining potential is
\begin{equation}
	\begin{split}
		V &=
		\mu_1^2|H_1|^2
		+\mu_2^2|H_2|^2
		+\frac{\lambda_1}{2}|H_1|^4
		+\frac{\lambda_2}{2}|H_2|^4 \\
		&\quad
		+\lambda_3|H_1|^2|H_2|^2
		+\lambda_4|H_1^\dagger H_2|^2 
		+\left[\frac{\lambda_5}{2}(H_1^\dagger H_2)^2+\text{h.c.}\right] \\
		&\quad
		+\mu_S^2|S|^2
		+\frac{\lambda_S}{2}|S|^4
		+\left[\frac{\mu_{sb}^2}{2}S^2+\text{h.c.}\right].
	\end{split}
	\label{eq:scalar_potential}
\end{equation}

The soft term $\mu_{sb}^2$ is the only source of explicit $U(1)_S$ breaking and gives the pNGB its mass \cite{Farhan:2026sym}:
\begin{equation}
	m_\phi^2=2|\mu_{sb}^2| .
	\label{eq:mphi_mass}
\end{equation}
This is technically natural because any radiative correction to $m_\phi^2$ must be proportional to $\mu_{sb}^2$ itself.

The spontaneous breaking of the global $U(1)_S$ symmetry at the scale $v_s$ would, in the absence of dilution, produce a network of global cosmic strings. For a global string, the energy per unit length scales as $\mu \sim 2\pi v_s^2 \ln(R/\delta)$, so the dimensionless tension is parametrically
\begin{equation}
	G\mu \;\sim\; \left(\frac{v_s}{M_{\rm Pl}}\right)^2,
	\label{eq:Gmu_estimate}
\end{equation}
up to logarithmic corrections. Current CMB bounds on a Nambu--Goto string network constrain $G\mu \lesssim 10^{-7}$ at 95\% CL, with stronger bounds $G\mu \lesssim 10^{-10}$ for global strings due to the long-range Goldstone boson fields they source~\cite{PlanckStringBounds,GlobalStringBounds}.

For a pNGB dark energy field of the type considered here, the pNGB decay constant is of order the Planck scale, $f_\phi \sim v_s \sim M_{\rm Pl}$, as required for the pNGB potential $V_0 \sim \mu_{sb}^2 v_s^2$ to reproduce the observed dark energy density with $m_\phi \sim H_0$. At such high symmetry-breaking scales, the cosmic string network formed at the $U(1)_S$ transition would be inconsistent with CMB bounds unless it is diluted. We therefore assume that the $U(1)_S$ breaking phase transition occurs before the end of inflation, so that any string network formed at that time is inflated away. This is the standard treatment for pNGB (axion-like) dark energy and dark matter models, in which the symmetry-breaking scale is typically large and the defect network would otherwise be cosmologically problematic~\cite{Marsh2016}. The assumption requires only that the reheating temperature after inflation be below $v_s$, so that the symmetry is not restored after reheating, which is automatic for $v_s \sim M_{\rm Pl}$. The EFT-Cutoff $\Lambda$ is inherited from \cite{Farhan:2026sym}.
	
	\subsection{Seesaw mechanism and radiative neutrino masses}
	\label{sec:seesaw}
	
	Neutrino masses are generated at one-loop level through the radiative seesaw mechanism. We introduce three right-handed neutrino singlets $N_{R_i}$ with a Majorana mass $M_N$ that is heavy compared to the electroweak scale. The neutrino sector Lagrangian is
	\begin{equation}
		\mathcal{L}_{\nu}
		=
		y_{ij} \bar{L}_{L_i} \tilde{H}_2 N_{R_j}
		+
		\frac{1}{2} M_{N_i} \bar{N}_{R_i}^c N_{R_i}
		+
		\text{h.c.},
		\label{eq:neutrino_lagrangian}
	\end{equation}
	where $L_{L_i}$ are the left-handed lepton doublets and $\tilde{H}_2 = i\sigma_2 H_2^*$ is the charge-conjugated inert doublet.
	
	Under the $Z_4$ symmetry, the right-handed neutrinos transform as $N_{R_i} \to -N_{R_i}$ (charge 2), while $H_2$ also transforms as $H_2 \to -H_2$; both fields are also odd under $Z_2^{\rm dark}$. This charge assignment allows the seesaw Yukawa term $\bar{L}_L \tilde{H}_2 N_R$ while forbidding the tree-level Dirac mass term $\bar{L}_L H_1 N_R$ that would involve the Standard Model Higgs $H_1$. Neutrino masses are therefore generated at one-loop level through the exchange of the neutral components of the inert doublet, $H_2^0$ and $A^0$ (the CP-even and CP-odd states), in the standard scotogenic mechanism. This radiative seesaw naturally links the smallness of neutrino masses to the dark sector, as the loop involves the dark matter candidate $H_2^0$.
	
	The one-loop amplitude for the neutrino self-energy is
\begin{equation}
	-i \Sigma_{ij}(p) = \int \frac{d^4 q}{(2\pi)^4} \frac{(-i y_{ik})(-i y_{jk})}{(q^2 - m_{H_2^0}^2)} \frac{i M_{N_k}}{q^2 - M_{N_k}^2} - (H_2^0 \to A^0).
\end{equation}
	After Wick rotation and performing the loop integral, the light neutrino mass matrix is (See Appendix A)
\begin{equation}
	\begin{split}
		m_{\nu}^{ij}
		&=
		\sum_k \frac{y_{ik} y_{jk} M_{N_k}}{16\pi^2}
		\Bigg[
		\frac{m_{H_2^0}^2}{m_{H_2^0}^2 - M_{N_k}^2}
		\ln\left(\frac{m_{H_2^0}^2}{M_{N_k}^2}\right) \\
		&\quad
		- \frac{m_{A^0}^2}{m_{A^0}^2 - M_{N_k}^2}
		\ln\left(\frac{m_{A^0}^2}{M_{N_k}^2}\right)
		\Bigg],
	\end{split}
	\label{eq:neutrino_mass}
\end{equation}
	where the mass splitting between the CP-even and CP-odd neutral scalars is $m_{A^0}^2 - m_{H_2^0}^2 = \lambda_5 v^2$, with $\lambda_5$ the quartic coupling from the scalar potential and $v = 174$~GeV the electroweak VEV. In the limit $m_{H_2^0}^2, m_{A^0}^2 \ll M_{N_k}^2$, Eq.~\eqref{eq:neutrino_mass} reduces to the standard scotogenic form, proportional to $\lambda_5 v^2$.
	
	A parameter scan of the scotogenic model establishes a linear relation between the absolute neutrino mass scale and the dark sector--Higgs coupling \cite{deBoer:2020Absolute}:
	\begin{equation}
		\lambda_5 = 3.1 \times 10^{-9} \left( \frac{m_{\nu_e}}{\text{eV}} \right),
		\label{eq:lambda5_nu_relation}
	\end{equation}
	which is independent of the neutrino mass hierarchy and CP phase \cite{deBoer:2020Absolute}. This relation provides a consistency condition linking the dark sector to the absolute neutrino mass scale. While \(\lambda_5\) does not explicitly enter the cosmological perturbation equations (as dark matter is treated as a cold fluid), it dictates the required mass splitting in the inert doublet for any chosen \(m_\nu\).  In our scan over the neutrino mass in Sec.~\ref{sec:results-sigma8}, Eq.~\eqref{eq:lambda5_nu_relation} therefore maps each curve to a specific, viable UV completion of the model. For the rest of the analysis, we set $m_\nu=0.06$ eV, which corresponds to $\lambda_5 \simeq 1.9 \times 10^{-10}$.
	
	Under the pNGB protection condition, the singlet sector is decoupled from the inert doublet, so $M_\rho$ and $M_\phi$ do not enter the neutrino mass loop. This is the standard scotogenic result. It is technically natural because setting $M_N \to 0$ restores lepton number symmetry.
	
	\subsection{Leading derivative couplings and neutrino interactions}
	\label{sec:interactions}
	
    With the renormalizable portals absent, the leading dark sector interactions are derivative operators consistent with all symmetries. The DM--DE interaction is generated at dimension six in the UV, with all couplings suppressed by the cutoff scale $\Lambda^2$~\cite{Farhan:2026sym}. We introduce three such couplings.
	
\subsubsection{Dark matter--dark energy momentum exchange}
\label{sec:dmde}

The DM--DE interaction is constructed from the $U(1)_S$ Noether current
\begin{equation}
	J_S^\mu = i\left(S^\dagger \partial^\mu S - (\partial^\mu S^\dagger) S\right).
	\label{eq:noether_current}
\end{equation}
After spontaneous symmetry breaking, with $S = (v_s + \rho) e^{i\phi/v_s}/\sqrt{2}$, the current becomes
\begin{equation}
	J_S^\mu = -\frac{(v_s + \rho)^2}{v_s} \partial^\mu \phi.
	\label{eq:noether_broken}
\end{equation}

We introduce the interaction
\begin{equation}
	\mathcal{L}_{\rm DM-DE}
	=
	\frac{c_6}{\Lambda^2}
	J_S^\mu
	\partial_\mu |H_2|^2,
	\label{eq:dmde_current}
\end{equation}
which is manifestly shift-symmetric under $\phi \to \phi + c$ and $U(1)_S$-invariant. Substituting Eq.~(\ref{eq:noether_broken}) and integrating out the heavy radial mode $\rho$ yields the effective low-energy operator
\begin{equation}
	\mathcal{L}_{\rm DM-DE}^{\rm eff}
	=
	\frac{c_6 v_s}{\Lambda^2}
	\partial^\mu \phi
	\partial_\mu |H_2|^2,
	\label{eq:dmde_eff_lagrangian}
\end{equation}

This generates a momentum-exchange rate
\begin{equation}
	\Gamma_{\rm DM}(a)
	=
	\xi\,H(a)\left[\frac{\rho_{\rm DM}(a)}{\rho_{\rm DM,0}}\right],
	\label{eq:dmde_gamma}
\end{equation}
as implemented in literature with a focus on dark matter relic density and $S_8$ tension prospects~\cite{Farhan:2026sym}.
	
	\subsubsection{Neutrino--dark energy derivative coupling}
	\label{sec:nude}
	
	We add a shift-symmetric derivative coupling between the pNGB field and the perturbation of the neutrino energy current,
\begin{equation}
	\mathcal{L}_{\nu-{\rm DE}}
	=
	\frac{g_{\phi\nu}}{\Lambda^2}
	\,
	\partial_\mu \phi
	\,
	\delta J^\mu_{\nu},
	\qquad
	\delta J^\mu_{\nu} \equiv J^\mu_{\nu} - \bar J^\mu_{\nu},
	\label{eq:nude_lagrangian}
\end{equation}
    where $J^\mu_{\nu} = (\rho_\nu + P_\nu)u^\mu_\nu$ is the neutrino energy current and $\bar J^\mu_{\nu} = (\bar\rho_\nu + \bar P_\nu)\bar u^\mu_\nu$ is its background value. The subtraction of the background current is a renormalization condition: the unsubtracted operator $\partial_\mu\phi\,J^\mu_\nu$ does not vanish on a homogeneous FRW background, because $\bar\phi' \neq 0$ and $\bar J^0_\nu = (\bar\rho_\nu + \bar P_\nu)/a \neq 0$. It would therefore mediate background energy exchange between the pNGB and the neutrino fluid. Any residual background transfer is absorbed into a redefinition of $\bar\rho_\nu$, $\bar P_\nu$, and the expansion rate $H(z)$, following the standard treatment of coupled-fluid interactions in cosmology~\cite{Wetterich1995,Amendola2000,Pourtsidou2013}. 

The subtracted operator vanishes on the background by construction, so the interaction acts as a pure momentum exchange at the level of perturbations, consistent with the scope of this work. More generally, a coupling mediates pure momentum exchange if and only if its background value vanishes identically. This operator is invariant under the shift symmetry $\phi \to \phi + c$ and does not generate a tree-level neutrino mass, unlike the Standard Model Yukawa term, which is forbidden by our $Z_4 \times Z_2^{\text{dark}}$ charge assignment. It preserves all symmetries and does not destabilize the pNGB mass, as it contains only derivative couplings of $\phi$. 
	
	\subsubsection{Dark matter--neutrino momentum exchange}
	\label{sec:dmn}
	
	We introduce a derivative coupling between the inert doublet $H_2$ and the perturbation of the neutrino energy current:
\begin{equation}
	\mathcal{L}_{\rm DM-\nu}
	=
	\frac{g_{\chi\nu}}{\Lambda^2}
	\,
	\partial_\mu |H_2|^2
	\,
	\delta J^\mu_{\nu},
	\label{eq:dmn_lagrangian}
\end{equation}
where $\delta J^\mu_\nu$ is defined in Eq.~\eqref{eq:nude_lagrangian} and $|H_2|^2$ is the gauge-invariant combination that contains the dark matter candidate $H_2^0$. This operator preserves the $Z_2^{\rm dark}$ symmetry since $|H_2|^2$ is invariant under $H_2 \to -H_2$, and does not generate a tree-level neutrino mass. The subtraction of the background current is not strictly required here, because $\partial_\mu|\bar H_2|^2 = 0$ on FRW; we adopt it for notational uniformity with the $\nu$--DE operator.
	
	\subsection{Full action and naturalness}
	\label{sec:action}
	
	The complete dark sector action is
	
	\begin{equation}
		\begin{split}
			\mathcal{S}_{\rm dark}
			&=
			\int d^4x \sqrt{-g}
			\Bigg[
			\frac{1}{2} g^{\mu\nu}\partial_\mu\phi\partial_\nu\phi
			- V(\phi)
			\\
			&\quad
			-\rho_{\rm DM}(n_{\rm DM})
			+ \mathcal{L}_{\rm DM-DE}
			+ \mathcal{L}_{\nu-{\rm DE}}
			+ \mathcal{L}_{\rm DM-\nu}
			\Bigg],
		\end{split}
		\label{eq:master_action}
	\end{equation}
	where $V(\phi)$ is the pNGB potential derived from Eq.~(\ref{eq:scalar_potential}) after spontaneous symmetry breaking, $\rho_{\rm DM}(n_{\rm DM})$ denotes the dark matter fluid energy density in the Schutz-Sorkin formalism, and the interaction terms are given in Eqs.~(\ref{eq:dmde_eff_lagrangian})--(\ref{eq:dmn_lagrangian}). The neutrino sector Lagrangian is given in Eq.~(\ref{eq:neutrino_lagrangian}), and the full scalar potential in Eq.~(\ref{eq:scalar_potential}). The Standard Model Lagrangian is implied but not shown for brevity.
	
All three interactions are technically natural:
\begin{itemize}
	\item The pNGB mass $m_\phi$ is protected by the $U(1)_S$ shift symmetry, which is only softly broken by $\mu_{sb}^2$. The derivative couplings of $\phi$ do not generate large radiative corrections to $m_\phi^2$.
	\item The dark matter mass $m_\chi$ is protected by the exact $Z_2^{\rm dark}$ symmetry. The leading operator that would destabilize $H_2^0$, namely $\mu_{12}^2 H_1^\dagger H_2$, is forbidden by $Z_2^{\rm dark}$ and cannot be generated at any loop order.
	\item The neutrino mass is generated radiatively through the seesaw mechanism and is protected by the approximate lepton number symmetry. The Majorana mass term $M_N$ is technically natural because setting $M_N \to 0$ restores lepton number symmetry.
\end{itemize}

Thus the model provides a unified dark sector with symmetry-protected dark matter, dark energy, and neutrino masses, all interacting through derivative momentum-exchange channels. The Class I singlet--doublet portals, which would mediate energy exchange between dark matter and dark energy, are set to zero as a model-building restriction that isolates the pure momentum-exchange limit.
	
	\section{Background and Perturbation Equations}
	\label{sec:background_perturbations}
	
	The background evolution of the unified dark sector is identical to the standard $Z_4 \times Z_2^{\rm dark}$-IDSM framework because all three interactions vanish at the level of the homogeneous background. The DM--DE and DM--$\nu$ operators contain derivatives of fields whose background values are constant ($\partial_\mu |\bar H_2|^2 = 0$ and $\partial_\mu |\bar\chi|^2 = 0$), and the $\nu$--DE operator is defined in terms of the subtracted current $\delta J^\mu_\nu$ [Eq.~\eqref{eq:nude_lagrangian}], which vanishes identically on FRW. Consequently, the background continuity equations for $\rho_{\rm DM}$ and $\rho_\nu$ retain their standard forms, and the only dynamical background component beyond $\Lambda$CDM is dark energy, which we model as a perfect fluid with constant equation of state $w_{\rm DE}=-0.99$, approximating the slow-roll regime of the pNGB field.
	
	\subsection{Background equations}
	
	The Friedmann equation is
	\begin{equation}
		H^2 = \frac{8\pi G}{3}
		\left(
		\rho_b + \rho_r + \rho_{\rm DM} + \rho_\nu + \rho_{\rm DE}
		\right),
	\end{equation}
	where $\rho_b$, $\rho_r$, $\rho_{\rm DM}$, $\rho_\nu$, and $\rho_{\rm DE}$ are the energy densities of baryons, radiation, cold dark matter, massive neutrinos, and dark energy, respectively.
	
	Baryons and radiation evolve according to
	\begin{equation}
		\rho_b' + 3\mathcal{H}\rho_b = 0,
	\end{equation}
	and
	\begin{equation}
		\rho_r' + 4\mathcal{H}\rho_r = 0.
	\end{equation}
	
	The cold dark matter density satisfies
	\begin{equation}
		\rho_{\rm DM}' + 3\mathcal{H}\rho_{\rm DM} = 0,
	\end{equation}
	so that $\rho_{\rm DM}\propto a^{-3}$. The momentum-exchange interactions do not alter this background scaling.
	
	For massive neutrinos, the background energy density and pressure are computed by the standard Fermi-Dirac distribution in \texttt{CLASS}. In the early universe, they behave like radiation, while at late times they transition to non-relativistic matter. Their background continuity equation is
	\begin{equation}
		\rho_\nu' + 3\mathcal{H}\left(1+w_\nu(a)\right)\rho_\nu = 0,
	\end{equation}
	where $w_\nu(a)=p_\nu/\rho_\nu$ is the time-dependent equation of state.
	
	For the dark energy component, we adopt the fluid description
	\begin{equation}
		\rho_{\rm DE}' + 3\mathcal{H}(1+w_{\rm DE})\rho_{\rm DE} = 0,
	\end{equation}
	with $w_{\rm DE}=-0.99$ and sound speed $c_s^2=1$. This choice approximates the slow-roll regime of the pNGB field, in which $w_{\rm DE}$ is very close to $-1$ and the sound speed is luminal, as appropriate for a canonical scalar field.
	
	\subsection{Linear perturbation equations}
	
	We work in the Newtonian (longitudinal) gauge \cite{Ma1995}. The perturbed line element is
	\begin{equation}
		ds^2 = a^2(\tau)\left[-(1+2\Psi)d\tau^2 + (1-2\Phi)\delta_{ij}dx^i dx^j\right],
	\end{equation}
	where $\Psi$ and $\Phi$ are the Bardeen potentials. Primes denote conformal time derivatives.
	
	The seesaw sector does not contribute to the perturbation equations at linear order because the heavy right-handed neutrinos $N_R$ decouple from the dark sector and the Standard Model at low energies. Their only effect is the generation of neutrino masses, which enter through the neutrino mass $m_\nu$ in the background and perturbation equations.
	
	The cold dark matter density contrast $\delta_{\rm DM}\equiv\delta\rho_{\rm DM}/\rho_{\rm DM}$ and velocity divergence $\theta_{\rm DM}\equiv\partial_i v_{\rm DM}^i$ obey
	\begin{equation}
		\delta_{\rm DM}' + \theta_{\rm DM} - 3\Phi' = 0,
	\end{equation}
	and the modified Euler equation
	\begin{equation}
		\theta_{\rm DM}' + \mathcal{H}\theta_{\rm DM}
		+ k^2\Psi
		=
		a\Gamma_{\rm DM-DE}(\theta_{\rm DE}-\theta_{\rm DM})
		+ a\Gamma_{\rm DM-\nu}(\theta_\nu-\theta_{\rm DM}).
	\end{equation}
	
	In the numerical implementation, dark energy is described as a fluid with $w=-0.99$ and $c_s^2=1$. The density contrast $\delta_{\rm DE}$ and velocity divergence $\theta_{\rm DE}$ obey
	\begin{equation}
		\delta_{\rm DE}' + (1+w)\left(\theta_{\rm DE} - 3\Phi'\right)
		+ 3\mathcal{H}(c_s^2-w)\delta_{\rm DE} = 0,
	\end{equation}
	and
	\begin{equation}
	\begin{split}
		\theta_{\rm DE}' &+ \mathcal{H}(1-3c_s^2)\theta_{\rm DE}
		- \frac{c_s^2}{1+w}k^2\delta_{\rm DE}
		+ k^2\Psi \\
		&= - a\frac{\rho_{\rm DM}}{\rho_{\rm DE}(1+w)}\Gamma_{\rm DM-DE}(\theta_{\rm DE}-\theta_{\rm DM}) \\
		&\quad - a\frac{\rho_\nu+p_\nu}{\rho_{\rm DE}(1+w)}\Gamma_{\nu-\rm DE}(\theta_{\rm DE}-\theta_\nu).
	\end{split}
\end{equation}

	For massive neutrinos, we implement the full Boltzmann hierarchy in \texttt{CLASS}, adding the drag terms only to the first ($\ell=1$) moment and leaving higher moments untouched. For analytical transparency we display the fluid-form first moment:
	\begin{equation}
	\begin{split}
		\theta_\nu' + \mathcal{H}(1 - 3w_\nu)\theta_\nu
		&- \frac{c_{s,\nu}^2}{1 + w_\nu} k^2 \delta_\nu
		+ k^2\Psi \\
		&=
		+ a\Gamma_{\nu-\rm DE}(\theta_{\rm DE}-\theta_\nu) \\
		&\quad
		- a\frac{\rho_{\rm DM}}{\rho_\nu+p_\nu}\Gamma_{\rm DM-\nu}(\theta_\nu-\theta_{\rm DM}),
	\end{split}
	\label{eq:dm_nu_drag}
    \end{equation}
	
	where $c_{s,\nu}^2 = \partial P_\nu / \partial \rho_\nu$ is the adiabatic sound speed of the neutrino fluid, determined by the Fermi-Dirac distribution.
	
	The momentum-exchange rates involving neutrinos are derived from the covariant fluid formalism (see Appendix~\ref{app:derivation}), whereas the DM-DE interaction parameterization takes the form \cite{Farhan:2026sym}:
	
	\begin{equation}
		\Gamma_{\rm DM-DE}(a)
		=
		\xi_{\rm DM-DE}\,H(a)\left[\frac{\rho_{\rm DM}(a)}{\rho_{\rm DM,0}}\right],
	\end{equation}
	
 	For neutrinos, the momentum-exchange rates obtained from the covariant drag structure described in Sec.~\ref{sec:derivation} are:
 \begin{equation}
 	\Gamma_{\nu-\rm DE}(a)
 	=
 	\frac{g_{\phi\nu}^2}{\Lambda^4} (\rho_\nu + P_\nu) H(a),
 	\label{eq:Gamma_nu_DE_main}
 \end{equation}
 and
 \begin{equation}
 	\Gamma_{\rm DM-\nu}(a)
 	=
 	\frac{g_{\chi\nu}^2}{\Lambda^4} (\rho_\nu + P_\nu) H(a).
 \end{equation}

    Here $g_{\phi\nu}$ and $g_{\chi\nu}$ are the effective dimensionless couplings used in the numerical implementation, and the $O(1)$ transport coefficients $C_{\phi\nu}, C_{\chi\nu}$ introduced in Sec.~\ref{sec:derivation} have been absorbed into their definitions. $\xi_{\rm DM-DE}$ is the effective DM--DE coupling ~\cite{Farhan:2026sym}, and $\Lambda$ is the EFT cutoff scale. Our numerical implementation extends this study to include dimensionless parameters $\xi_{\rm DM-\nu}$ and $\xi_{\nu-\rm DE}$ which incorporate the neutrino interactions:
    
    \begin{equation}
    	\xi_{\rm DM-\nu}
    	=
    	\frac{\Gamma_{\rm DM-\nu}(a)}{H(a)}
    	=
    	\frac{g_{\chi\nu}^2}{\Lambda^4} (\rho_\nu + P_\nu).
    	\label{eq:keynu-DM}
    \end{equation}
    
        \begin{equation}
    	\xi_{\rm \nu-DE}
    	=
    	\frac{\Gamma_{\rm DE-\nu}(a)}{H(a)}
    	=
    	\frac{g_{\phi\nu}^2}{\Lambda^4} (\rho_\nu + P_\nu).
    	\label{eq:keynu-DE}
    \end{equation}
	
	For the analytical derivation in this paper, we use the fluid approximation for neutrinos, treating them as a perfect fluid with equation of state $w_\nu(a)$ and sound speed $c_{s,\nu}^2(a)$. This captures the essential physics of the momentum-exchange couplings. In the \texttt{CLASS} implementation, the full neutrino Boltzmann hierarchy is employed, with the drag terms entering only at the first moment.

\subsection{Covariant derivation of the neutrino momentum-transfer structure}
\label{sec:derivation}

The neutrino drag rates are derived and parameterized from the perfect-fluid stress-energy tensor, as shown in Appendix~\ref{app:derivation}. The covariant conservation equation in the presence of an interaction is
\begin{equation}
	\nabla_\mu T^{\mu\nu} = Q^\nu,
	\label{eq:conservation_sec}
\end{equation}
where $T^{\mu\nu}$ is the neutrino stress-energy tensor and $Q^\nu$ is the four-force density acting on the neutrino fluid. We impose pure momentum exchange by requiring $u_\mu Q^\mu = 0$, which guarantees no energy transfer in the local neutrino rest frame.

For an isotropic drag interaction between neutrinos and another component $X$, covariance and linearity in the relative velocity imply the leading-order four-force
\begin{equation}
	Q^\mu
	=
	-(\rho_\nu + P_\nu)\Gamma_{\nu X}
	\Delta^\mu{}_\alpha
	(u^\alpha - u_X^\alpha),
	\label{eq:Q_general_sec}
\end{equation}
where $\Delta^{\mu\alpha} = g^{\mu\alpha} + u^\mu u^\alpha$ is the projector onto the neutrino rest frame, and $\Gamma_{\nu X}$ is the momentum-transfer rate.

The enthalpy prefactor $\rho_\nu + P_\nu$ in Eq.~\eqref{eq:Q_general_sec} is not a normalization choice. Insisting that the transport coefficient $\Gamma_{\nu X}$ be a genuine rate --- with units of inverse time, independent of the fluid state, and not a combination of fluid variables that happens to have those units in one particular regime --- forces the four-force density to be proportional to the momentum density $(\rho_\nu + P_\nu)u^\mu$ rather than to $\rho_\nu u^\mu$. Absorbing the prefactor into a redefined rate $\tilde\Gamma_{\nu X} \equiv (\rho_\nu+P_\nu)\Gamma_{\nu X}$ would make $\tilde\Gamma_{\nu X}$ a function of the fluid state, which defeats the purpose of matching to the microscopic operator in Sec.~\ref{sec:transport}. In Newtonian gauge, the spatial projection gives the Euler equation
\begin{equation}
	\theta_\nu'
	+
	\mathcal H(1-3w_\nu)\theta_\nu
	-
	\frac{c_{s,\nu}^2}{1 + w_\nu}k^2\delta_\nu
	+
	k^2\Psi
	=
	-a\Gamma_{\nu X}
	(\theta_\nu-\theta_X).
	\label{eq:nu_euler_drag_sec}
\end{equation}

The remaining quantity $\Gamma_{\nu X}$ is a transport coefficient. Dimensional analysis of the dimension-six interaction gives
\begin{equation}
	\Gamma_{\nu X}
	\sim
	C_{\nu X}
	\frac{g_{\nu X}^2}{\Lambda^4}
	(\rho_\nu + P_\nu)
	\mathcal E_X,
	\label{eq:general_rate_sec}
\end{equation}
where $\mathcal E_X$ is the characteristic inverse timescale and $C_{\nu X}$ is a dimensionless coefficient. For the cosmological implementation, we adopt the adiabatic closure $\mathcal E_X = H(a)$, yielding
\begin{equation}
	\Gamma_{\nu-\rm DE}(a)
	=
	C_{\phi\nu}
	\frac{g_{\phi\nu}^2}{\Lambda^4}
	(\rho_\nu + P_\nu)H(a),
	\label{eq:Gamma_nu_DE_derived_sec}
\end{equation}
and
\begin{equation}
	\Gamma_{\rm DM-\nu}(a)
	=
	C_{\chi\nu}
	\frac{g_{\chi\nu}^2}{\Lambda^4}
	(\rho_\nu + P_\nu)H(a).
	\label{eq:Gamma_DM_nu_derived_sec}
\end{equation}

The coefficients $C_{\phi\nu}$ and $C_{\chi\nu}$ encode short-distance transport details and are absorbed into the effective couplings used in the numerical analysis. A robust consequence of the covariant drag structure is the enthalpy weighting $\Gamma_{\nu X} \propto \rho_\nu + P_\nu$. For relativistic neutrinos, $P_\nu = \rho_\nu/3$, giving a $4/3$ enhancement relative to a pressureless fluid at fixed energy density. The full derivation is given in Appendix~\ref{app:derivation}.

\section{Stability Conditions}
\label{sec:stability}

The unified dark sector model introduces three momentum-exchange couplings. For the linear perturbation theory to be well-defined and numerically tractable, the following conditions must be satisfied.

\subsection{No-ghost condition}

In our numerical implementation, dark energy is treated as a perfect fluid with a constant equation of state $w_0 = -0.99$ and sound speed $c_s^2 = 1$. This serves as a reliable approximation to the slow-roll regime of the pNGB field $\phi$, whose canonical kinetic term is positive definite. The fluid description does not introduce ghost degrees of freedom: for $w > -1$, the fluid is stable, and the choice $c_s^2 = 1$ matches the sound speed of a canonical scalar field in its rest frame.

Cold dark matter is pressureless, and its kinetic term is positive for $\rho_{\rm DM}>0$. Massive neutrinos are treated through the standard Boltzmann hierarchy, which is free from ghost pathologies for physical neutrino masses. The new neutrino interactions are derivative couplings that do not modify the kinetic terms of the neutrino fields, so they do not introduce ghosts.

For dark energy, $\rho_{\mathrm{DE}} + P_{\mathrm{DE}}$ is small but non-zero for $w_0 = -0.99$. The back-reaction terms in the dark energy Euler equation are proportional to $1/(\rho_{\mathrm{DE}} + P_{\mathrm{DE}})$, which can be large. This does not imply self-interaction; rather, it reflects the fact that the dark energy fluid is sensitive to the enthalpy density of other fluids. The dark energy field itself does not have a self-drag term because the interaction couples it to other fluids, not to itself.

\subsection{Gradient stability}

In the quasi-static limit, the propagation speeds of the coupled system determine whether gradient instabilities occur. For the dark energy fluid with sound speed $c_s^2=1$, the propagation speed is luminal:
\begin{equation}
	c_{\rm DE}^2 = c_s^2 = 1 .
\end{equation}
The dark matter fluid has $c_s^2=0$ as usual. The neutrino fluid has a time-dependent adiabatic sound speed $c_{s,\nu}^2(a)$ determined by the standard Fermi-Dirac distribution,
\begin{equation}
	c_{s,\nu}^2 = \frac{\partial P_\nu}{\partial \rho_\nu},
\end{equation}
which is always positive.

The momentum-exchange terms do not introduce additional spatial-gradient terms in the Euler equations. They act as velocity-drag terms and therefore do not alter the sound speeds of the individual components. Consequently, within the fluid/EFT truncation adopted here, we find no gradient instability in the parameter region considered.

\subsection{Strong-coupling limit and tight-coupling approximation}

When a momentum-exchange rate becomes much larger than the Hubble rate, the corresponding relative velocity is driven to zero on a timescale much shorter than the Hubble time. In this regime, the numerical integration of the perturbation equations becomes stiff, but no physical instability occurs.

To handle this, we implement a momentum-conserving cap on the drag rates, analogous in spirit to the tight-coupling approximation used for the photon--baryon system but with a different regularization scheme. For each interaction, when the \emph{larger} of the two rates that define an interacting pair exceeds $20\,\mathcal{H}$, we replace both rates by a common factor that scales the pair down to the threshold while preserving their ratio. The velocities do not track each other exactly in this regime --- rather, the drag is capped at a finite effective rate $\sim 20\,\mathcal{H}$ on both sides of the pair. The cap is applied when
\begin{equation}
	\max\!\left(a\,\Gamma_{\rm DM-DE},\,a\,\Gamma_{\rm DM-DE}\,R_{\rm fld}\right) > 20\,\mathcal{H},
\end{equation}
while the neutrino--DE drag is dropped when
\begin{equation}
	\max\!\left(a\,\Gamma_{\nu-\rm DE},\,a\,\Gamma_{\nu-\rm DE}\,R_{\nu\text{-DE}}\right) > 20\,\mathcal{H},
\end{equation}
and similarly for the DM--neutrino drag with reference rate $\max(a\Gamma_{\rm DM-\nu}, a\Gamma_{\rm DM-\nu} R_{\nu\text{-DM}})$. The cap is applied to the \emph{maximum} of the two rates that define a given interacting pair, and the same scale factor is applied to both sides of the pair. This guarantees that the ratio $\rho_A\Gamma_A / \rho_B\Gamma_B$ is preserved by the cap, so total momentum conservation holds exactly in the strong-coupling regime. Outside these regimes, the full coupled equations are solved exactly. We have verified that our results are insensitive to the precise value of the threshold in the DM–DE case: varying the coefficient from $5$ to $50$ changes $\sigma_8$ by less than 0.1\%.

Because the neutrino rates scale with the enthalpy density $(\rho_\nu+P_\nu)$, which falls faster than the CDM density at late times, the neutrino couplings are dominated by the epoch when neutrinos transition from relativistic to non-relativistic, rather than by the drag rate at $z=0$. In the strong-coupling limit, the DM--$\nu$ channel nevertheless saturates at a finite value of $\sigma_8$, just as the DM--DE coupling does in the original model~\cite{Farhan:2026sym}. The saturation value is set by the integrated drag over the epoch of structure formation and not by the instantaneous rate at late times.

Total momentum conservation among the three interacting components is maintained by construction in the resolved regime, where each pairwise drag term in one Euler equation appears with opposite sign in the conjugate equation. In each tight-coupling regime, the corresponding pair of velocities is driven to equality, and the momentum exchange is determined by the small residual velocity difference. When DE is simultaneously tightly coupled to both DM and neutrinos, all three velocities are driven to a common value and the net momentum transfer is fixed by the small residual differences that enforce this equality.
	
\section{Methodology}
\label{sec:method}

We implement the unified dark sector model in the Boltzmann solver \texttt{CLASS}~\cite{Lesgourgues2011,Blas2011}. The background evolution is that of a flat $\Lambda$CDM cosmology with a dark energy component that we model as a perfect fluid with equation of state $w_{\rm DE}=-0.99$ and sound speed $c_s^2=1$. We fix these values as an approximation to the slow-roll regime of the pNGB dark energy field, which is extremely close to de Sitter but not exactly a cosmological constant. This simplification is well-justified because the pNGB is very close to $w=-1$ during the epoch relevant for structure formation, and the momentum-exchange effects are insensitive to the precise value of $w_{\rm DE}$ provided it is close to $-1$. The perturbation module is modified to include the three momentum-exchange interactions described below.

\subsection{Modifications to \texttt{CLASS}}

\subsubsection{Dark matter Euler equation}

The cold dark matter velocity divergence receives two drag terms:
\begin{equation}
	\theta_{\rm cdm}' + \mathcal{H}\theta_{\rm cdm}
	=
	a\Gamma_{\rm DM-DE}(\theta_{\rm DE}-\theta_{\rm cdm})
	+ a\Gamma_{\rm DM-\nu}(\theta_\nu-\theta_{\rm cdm}).
\end{equation}

\subsubsection{Dark energy Euler equation}

The dark energy velocity divergence receives backreaction terms from both the DM--DE and the $\nu$--DE interactions:
\begin{equation}
	\begin{split}
		&\theta_{\rm DE}' + \mathcal{H}(1-3c_s^2)\theta_{\rm DE}
		- \frac{c_s^2}{1+w} k^2 \delta_{\rm DE} \\
		&\quad
		=
		- a \frac{\rho_{\rm cdm}}{\rho_{\rm DE}(1+w)}\Gamma_{\rm DM-DE}(\theta_{\rm DE}-\theta_{\rm cdm}) \\
		&\quad
		- a \frac{\rho_\nu+P_\nu}{\rho_{\rm DE}(1+w)}\Gamma_{\nu-\rm DE}(\theta_{\rm DE}-\theta_\nu).
	\end{split}
\end{equation}

\subsubsection{Neutrino Boltzmann hierarchy}

For massive neutrinos, we solve the full Boltzmann hierarchy. The momentum-exchange interactions enter only through the first moment ($\ell=1$) for each momentum mode $q$; this corresponds to treating the interaction as a pure drag force, which is the leading-order effect in the fluid limit. Higher multipoles are suppressed by powers of $k/\Gamma$ in the tightly coupled regime and are not expected to contribute significantly to the observables considered here:
\begin{equation}
	\begin{split}
		\Psi_1'(q) \;\supset\;
		&-\,a\,\Gamma_{\nu-\rm DE}
		\Bigl(\Psi_1(q) - \frac{q\,\theta_{\rm DE}}{3k}\Bigr) \\
		&-\,a\,\Gamma_{\rm DM-\nu}\,
		\frac{\rho_{\rm DM}}{\rho_\nu+P_\nu}
		\Bigl(\Psi_1(q) - \frac{q\,\theta_{\rm cdm}}{3k}\Bigr),
	\end{split}
	\label{eq:hierarchy_drag}
\end{equation}
where $\Psi_1(q)$ is the first multipole of the neutrino phase-space perturbation, related to the neutrino velocity divergence by $\theta_\nu(q) \equiv 3k\Psi_1(q)/q$ in the CLASS convention. Higher moments ($\ell \ge 2$) remain untouched. The equilibrium values $q\,\theta_X/(3k)$ appearing inside the drag brackets are precisely those that make $\theta_\nu(q) \to \theta_X$ in the strong-coupling limit, and the density ratio $\rho_{\rm DM}/(\rho_\nu+P_\nu)$ in the DM--$\nu$ term is the same factor that appears on the right-hand side of Eq.~\eqref{eq:dm_nu_drag}; it enforces antisymmetry of the two drag terms under exchange of the interacting pair and ensures exact total momentum conservation. Because the drag term is defined at the level of the momentum-resolved hierarchy, we evolve neutrinos with the exact Boltzmann hierarchy throughout, setting \texttt{ncdm\_fluid\_approximation = 3} in the CLASS input file (the enum value \texttt{ncdmfa\_none}). CLASS parses this parameter as an integer and not as a string. This avoids a spurious switch-off of the drag at low redshift, where the fluid approximation would otherwise replace the hierarchy in which the interaction is formulated.

\subsection{Momentum-exchange rates}

For the DM--DE interaction we adopt the Type-3 parameterization \cite{Farhan:2026sym},
\begin{equation}
	\Gamma_{\rm DM-DE}(a)
	=
	\xi_{\rm DM-DE}\, H(a)\, \frac{\rho_{\rm cdm}(a)}{\rho_{\rm cdm,0}},
\end{equation}
where the coupling $\xi_{\rm DM-DE}$ absorbs the microphysical parameters $c_6$, $v_s$, $\Lambda_2$, and $m_{H_2}^2$ of the underlying $Z_4$-IDSM model.

The neutrino--dark energy and dark matter--neutrino rates are derived from the covariant fluid formalism discussed in Sec.~\ref{sec:derivation} and Appendix~\ref{app:derivation}. They are proportional to the enthalpy density $\rho_\nu+P_\nu$. In terms of the dimensionless couplings $g_{\phi\nu}$ and $g_{\chi\nu}$, they read
\begin{equation}
	\Gamma_{\nu-\rm DE}(a)
	=
	\frac{g_{\phi\nu}^2}{\Lambda^4}(\rho_\nu+P_\nu)H(a),
\end{equation}
\begin{equation}
	\Gamma_{\rm DM-\nu}(a)
	=
	\frac{g_{\chi\nu}^2}{\Lambda^4}(\rho_\nu+P_\nu)H(a).
\end{equation}
In the CLASS implementation, these couplings are named \texttt{g\_nu} and \texttt{g\_chinu}, respectively; the mapping between the two notations is one-to-one and appears nowhere else in the code (apart from their initialization in \texttt{input.c} and \texttt{perturbations.h}). While the cosmological observables depend only on the effective combination \(g^2/\Lambda^4\), for numerical evaluation we fix the EFT cutoff to \(\Lambda = 10^8\) GeV, consistent with the \(Z_4\)-IDSM framework~\cite{Farhan:2026sym}. 

In the relativistic regime ($P_\nu=\rho_\nu/3$), both rates are enhanced by a factor of $4/3$ relative to the non-relativistic limit. This enthalpy-weighted scaling is a distinctive consequence of the first-principles derivation and differs from phenomenological drag parameterizations in which the rate is simply proportional to $\rho_\nu$ or treated as a free function of scale factor.

\subsection{Tight-coupling approximation}

In the early universe, the dark energy density is negligible compared to the CDM and neutrino densities. As a result, the backreaction ratios appearing in the dark energy Euler equation,
\begin{equation}
	R_{\rm fld} \equiv \frac{\rho_{\rm cdm}}{\rho_{\rm DE}(1+w_{\rm DE})},
	\qquad
	R_{\nu\text{-DE}} \equiv \frac{\rho_\nu+P_\nu}{\rho_{\rm DE}(1+w_{\rm DE})},
\end{equation}
become very large, and the coupled system becomes numerically stiff.

To handle this regime without introducing arbitrary cutoffs that would violate momentum conservation, we implement a momentum-conserving cap on the drag rates, analogous in spirit to the tight-coupling approximation used for the photon--baryon system but with a different regularization scheme. For each interaction, when the \emph{larger} of the two rates that define an interacting pair exceeds $20\,\mathcal{H}$, we replace both rates by a common factor that scales the pair down to the threshold while preserving their ratio. The velocities do not track each other exactly in this regime; rather, the effective drag is capped at $\sim 20\,\mathcal{H}$ on both sides of the pair. Outside the capped regime, the full coupled equations are solved exactly.

We have verified that our results are insensitive to the precise value of the threshold in the DM--DE case: varying the coefficient from $5$ to $50$ changes $\sigma_8$ by less than $0.1\%$. The same insensitivity is expected for the neutrino interactions, because the cap only regulates the integrator in the regime where the pair of velocities is already locked and the physical drag rate is symmetric across the cap.

Once the dark energy density becomes non-negligible, the drag rates no longer exceed $20\mathcal{H}$, and the full coupled equations are solved with unregularized ratios. This ensures that total momentum conservation among the interacting components holds in the resolved regime, while in each tight-coupling regime the relevant velocities are nearly equal.

\section{Results}
\label{sec:results}

In this section we present the cosmological signatures of the three momentum-exchange channels of the unified dark sector. We structure the analysis as follows: Sec.~\ref{sec:results-sigma8} and Fig.~\ref{fig:FIG1} detail the non-monotonic clustering amplitude $\sigma_8$ and its additivity with the neutrino mass. Section~\ref{sec:results-cmb} and Figs.~\ref{fig:FIG3} and~\ref{fig:FIG2} show the primary CMB temperature and lensing potential ratios, highlighting the distinctive lensing enhancement of the DM--$\nu$ channel. Section~\ref{sec:results-fsigma8} and Fig.~\ref{fig:FIG4} compare the growth rate $f\sigma_8(z)$ against RSD data. Section~\ref{sec:results-pk} and Fig.~\ref{fig:FIG5} present the linear matter power spectrum and the severe small-scale constraint on the DM--$\nu$ coupling. Finally, Sec.~\ref{sec:results-enthalpy} and Fig.~\ref{fig:FIG6} illustrate the enthalpy weighting that distinguishes our covariant derivation from phenomenological drag parameterizations.

Unless otherwise stated, we work in a fiducial cosmology with $h = 0.6781$, $\omega_b = 0.02238$, $\omega_{\rm cdm} = 0.1201$, $N_{\rm ur} = 2.0308$, $N_{\rm ncdm} = 1$,
$m_\nu = 0.06$~eV, $w_{\rm DE} = -0.99$, $c_s^2 = 1$, $A_s = 2.1\times 10^{-9}$, $n_s = 0.966$, $\tau_{\rm reio} = 0.054$, and we fix $g_{\phi\nu} = 0$ except where the $\nu$--DE channel is explicitly discussed.

\subsection{Clustering amplitude and its degeneracy with the neutrino mass}
\label{sec:results-sigma8}

\begin{figure}[t]
	\centering
	\includegraphics[width=\columnwidth]{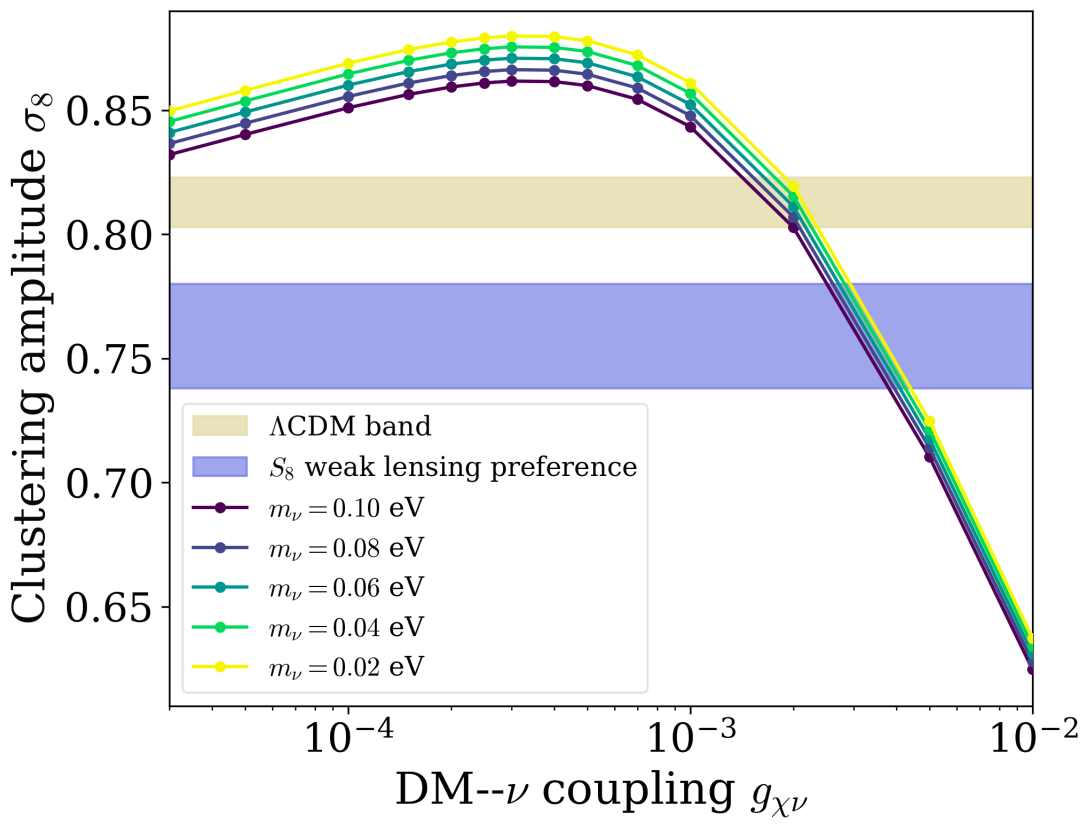}
	\caption{Linear clustering amplitude $\sigma_8$ as a function of the DM--$\nu$ coupling $g_{\chi\nu}$, for five values of the total neutrino mass. The tan band shows the neutrino-mass-dependent $\Lambda$CDM
	range; the blue band shows the $S_8$ preference from weak lensing. The five curves are parallel across the entire coupling range, demonstrating that the coupling and neutrino mass modify $\sigma_8$ independently. Since $g_{\chi\nu}$ is plotted logarithmically, the lowest value shown in the plot ($g_{\chi\nu}=3\times 10^{-5}$) lies outside the tan-colored baseline $\Lambda$CDM band.}
	\label{fig:FIG1}
\end{figure}

Figure~\ref{fig:FIG1} shows the linear clustering amplitude $\sigma_8$ as a function of the DM--$\nu$ coupling $g_{\chi\nu}$ for five values of the total neutrino mass, spanning $m_\nu = 0.02$~eV to $m_\nu = 0.10$~eV. The curves exhibit a clear non-monotonic structure. The baseline $g_{\chi\nu} = 0$ values sit inside the neutrino-mass-dependent $\Lambda$CDM band (tan shaded region), with $\sigma_8$ decreasing monotonically as the total neutrino mass increases, reaching $0.803$ at $m_\nu = 0.10$~eV. Since the figure starts with $g_{\chi\nu}=3 \times 10^{-5}$, the tan shaded $\Lambda$CDM band depicts the baseline values for reference. As the coupling increases, each curve rises monotonically, reaches a peak at $g_{\chi\nu} \simeq 3\times 10^{-4}$, and then collapses sharply at $g_{\chi\nu} \gtrsim 10^{-3}$. The peak amplitude for the fiducial $m_\nu = 0.06$~eV is $\sigma_8 = 0.871$, approximately $7.4\%$ above the corresponding $\Lambda$CDM value of $0.811$.

This non-monotonic behavior reflects a competition between late-time and early-time dynamics. At moderate couplings, the drag acts primarily at late times, damping neutrino free-streaming and allowing neutrinos to fall into the CDM potential wells, thereby boosting the total clustering amplitude. However, because the drag rate scales as \((1+z)^6\), a larger coupling forces the CDM and neutrinos into tight coupling at very high redshifts. In this strong-coupling regime, the CDM is dragged along by the relativistic, free-streaming neutrinos, which prevents the CDM from forming its own potential wells. This early-time suppression washes out the small-scale perturbations and ultimately dominates over the late-time enhancement, causing \(\sigma_8\) to collapse.

The most striking feature of Fig.~\ref{fig:FIG1} is the additivity of the neutrino-mass and coupling effects. The five curves are essentially parallel across the entire range of $g_{\chi\nu}$. The vertical offset between adjacent curves is $|\Delta\sigma_8| \simeq 0.020$--$0.021$, independent of the coupling strength, and it matches the $\Lambda$CDM neutrino-mass sensitivity to within a few percent. The location and width of the peak are unchanged as $m_\nu$ varies. Equivalently, in the
language of parameter constraints, the DM--$\nu$ coupling and the total neutrino mass are not degenerate with respect to $\sigma_8$: the two parameters span the $(\sigma_8, m_\nu)$ plane independently, allowing each to be constrained by the same data separately.

The peak value $g_{\chi\nu} \simeq 3\times 10^{-4}$ produces $\sigma_8 = 0.871$, which sits well above the $S_8$ weak-lensing preference (blue shaded region in Fig.~\ref{fig:FIG1}) derived from KiDS-1000~\cite{KiDS2021} and DES Year~3~\cite{DESY3}. To reconcile the Planck and weak-lensing amplitudes, the coupling must be increased further to \(g_{\chi\nu} \sim 4 \times 10^{-3}\), where the drag-induced suppression pushes \(\sigma_8\) into the blue band. However, as we show in Sec.~\ref{sec:results-pk}, both this \(S_8\)-resolving regime and the enhancing peak are excluded by small-scale structure. The viable parameter space at the $1\sigma$ level lies below \(g_{\chi\nu} \lesssim 10^{-5}\), where the effect on \(\sigma_8\) is negligible at best.

The two other channels do not exhibit comparable behaviour. The DM--DE coupling $\xi_{DM}$ produces a monotonic suppression of $\sigma_8$ that saturates at $-2\%$ for $\xi_{DM} \gtrsim 0.3$ (we do not show
this separately as it reproduces already established results~\cite{Farhan:2026sym}). The $\nu$--DE coupling $g_{\phi\nu}$ has a negligible effect on $\sigma_8$; even at saturation the shift is below $0.02\%$, consistent with the fact that this channel does not couple to CDM.

\subsection{Primary CMB and lensing potential}
\label{sec:results-cmb}

\begin{figure}[t]
	\centering
	\includegraphics[width=\columnwidth]{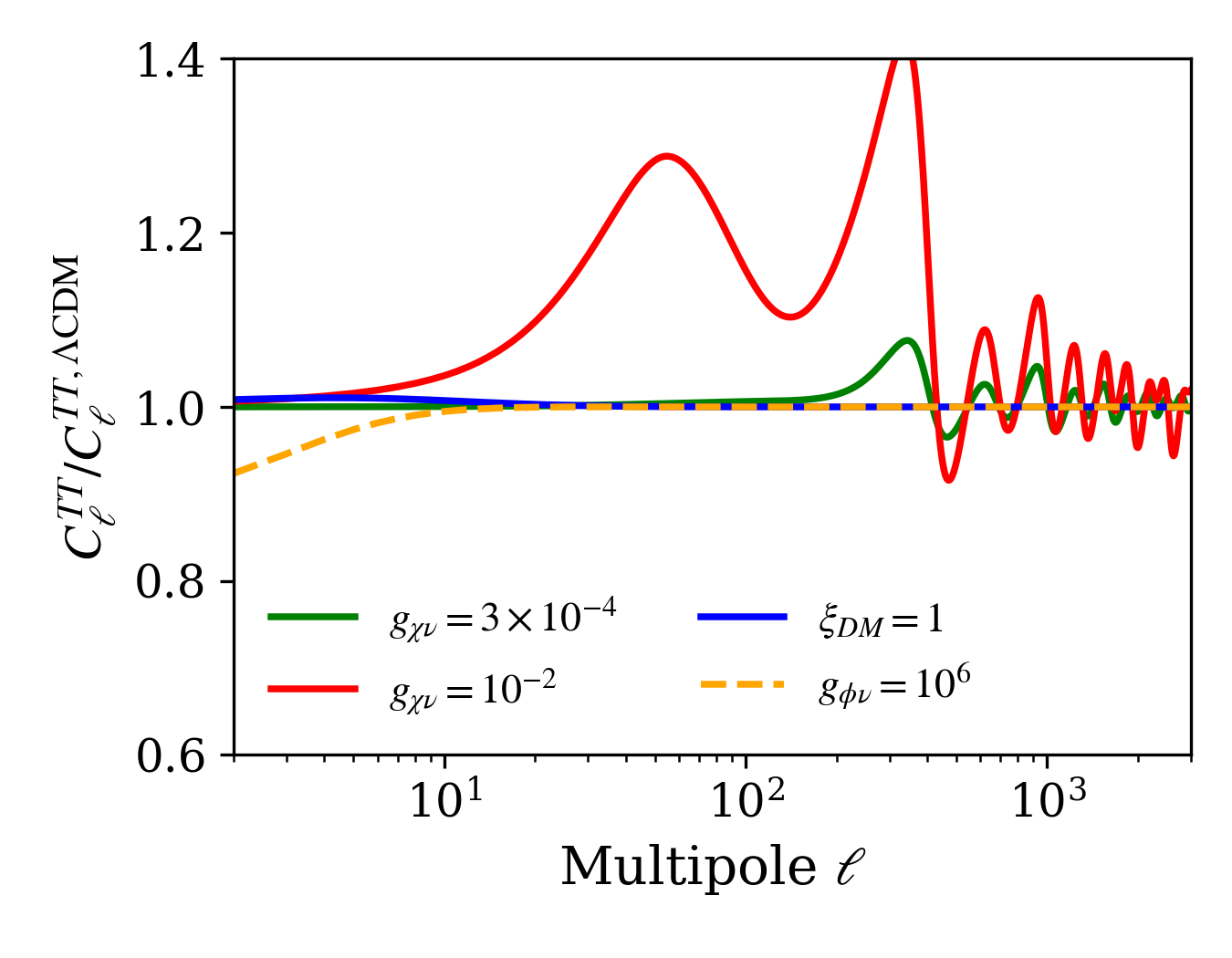}
	\caption{Primary CMB temperature ratio with respect to \(\Lambda\)CDM, for four representative couplings. The viable couplings (\(g_{\chi\nu}
	\le 3\times 10^{-4}\), \(\xi_{DM} = 1\), \(g_{\phi\nu} \le 10^6\)) leave the acoustic peaks virtually untouched. The extreme, excluded coupling
	\(g_{\chi\nu} = 10^{-2}\) (red) produces massive oscillatory deviations at \(\ell \gtrsim 200\), completely distorting the damping tail.}
	\label{fig:FIG3}
\end{figure}

\begin{figure}[t]
	\centering
	\includegraphics[width=\columnwidth]{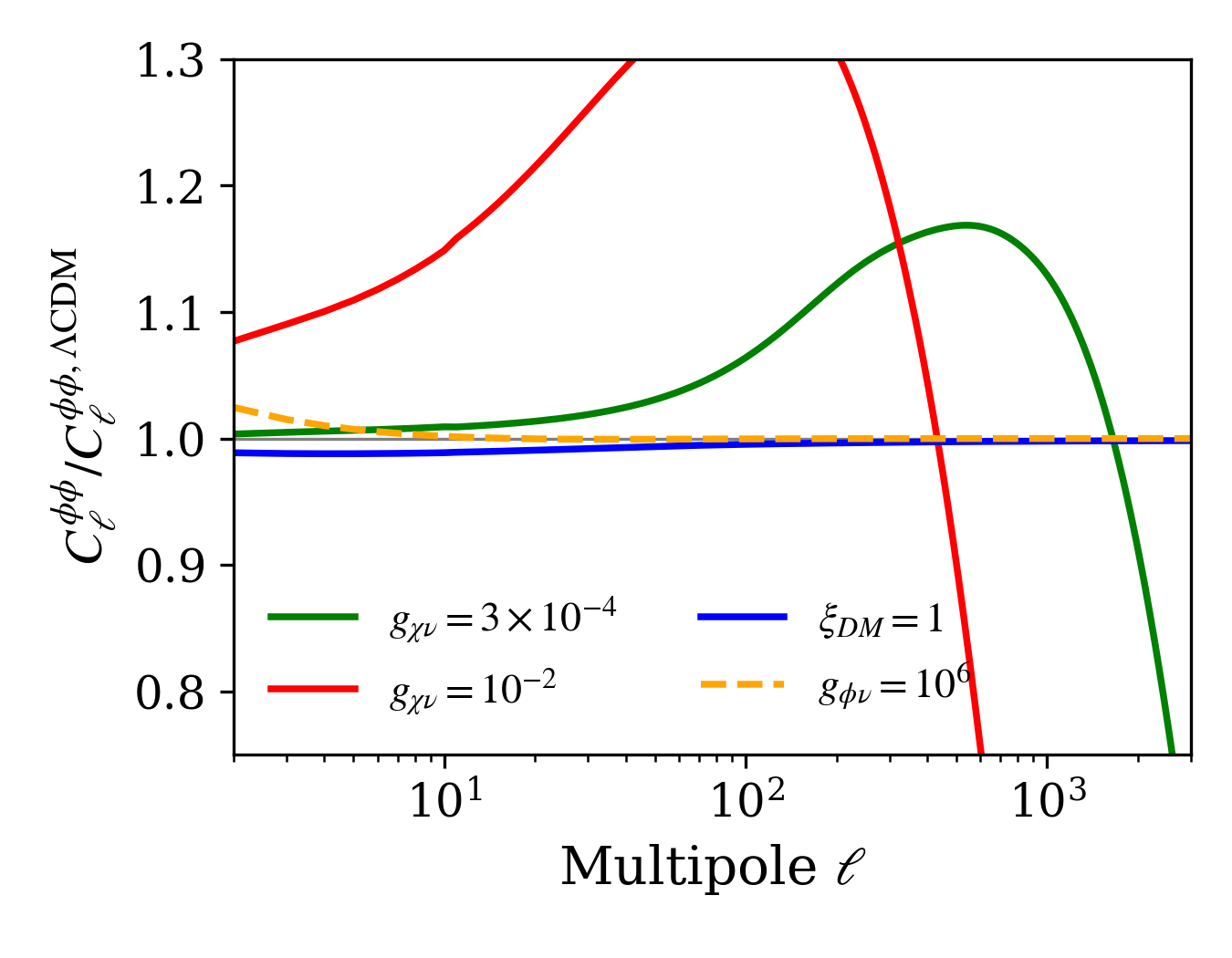}
	\caption{CMB lensing potential ratio with respect to $\Lambda$CDM. The DM--$\nu$ channel enhances the lensing potential, reaching $\sim 17\%$ at $\ell=500$ for $g_{\chi\nu}=3\times 10^{-4}$ and still $\sim 5\%$ at $\ell=1000$ for the viable bound. The $\nu$--DE and DM--DE channels are null or weakly suppressing.}
	\label{fig:FIG2}
\end{figure}

Figures~\ref{fig:FIG3} and~\ref{fig:FIG2} show the primary CMB temperature ratio and the CMB lensing potential ratio, respectively, for four representative parameter choices. These points are chosen to isolate distinct physical regimes: the maximum viable enhancement in the DM--\(\nu\) channel (\(g_{\chi\nu} = 3\times 10^{-4}\)), an extreme excluded saturation limit of the DM--\(\nu\) channel (\(g_{\chi\nu} = 10^{-2}\)), and the saturation limits of the DM--DE (\(\xi_{DM}=1\)) and \(\nu\)--DE (\(g_{\phi\nu}=10^6\)) channels. The ratios are computed with respect to the $\Lambda$CDM baseline.

In the primary CMB (Fig.~\ref{fig:FIG3}), the viable couplings leave the acoustic peaks unchanged at the percent level. The $\nu$--DE (orange) and DM--DE (blue) channels are visually indistinguishable from $\Lambda$CDM, though tiny deviations exist at $\ell \lesssim 10$ where cosmic variance is overwhelmingly large. The DM--$\nu$ coupling at the \(\sigma_8\)-peak value \(g_{\chi\nu} = 3\times 10^{-4}\) (green curve) produces a distinct oscillatory feature: a \(\sim 0.6\%\) excess at \(\ell \sim 100\), growing to a \(\sim 5\%\) peak near \(\ell \sim 300\), followed by a \(2.5\%\) dip near \(\ell \sim 500\). While these oscillations represent a testable signature of the model, they remain well inside the Planck 2018 \(1\sigma\) error bars. Only the excluded coupling $g_{\chi\nu} = 10^{-2}$ (red curve) produces massive $O(10-30\%)$ oscillatory deviations at $\ell \gtrsim 200$, catastrophically distorting the damping tail. At multipoles $\ell \lesssim 30$ cosmic variance dominates and no coupling leaves a detectable imprint. We therefore conclude that the primary CMB does not constrain the momentum-exchange couplings at viable values; its role in a full likelihood analysis will be to anchor the standard cosmological parameters against which the couplings are measured.

In the CMB lensing potential (Fig.~\ref{fig:FIG2}), the picture is qualitatively different. The lensing potential is an integrated quantity sensitive to the growth of the Weyl potential at $z \sim 0.5$--$3$, precisely the epoch where the drag is active. Three features stand out. First, the $\nu$--DE channel is null: even at saturation $g_{\phi\nu} = 10^{6}$ (orange), the $\phi\phi$ ratio is $1.0000$ to four decimal places across the entire high-multipole range ($\ell \gtrsim 20$). A tiny deviation from unity exists at the lowest multipoles, $\ell \sim 2-10$, but it is deeply buried in cosmic variance. This is expected, since the $\nu$--DE drag couples two subdominant species---neutrinos and dark energy---without touching the cold dark matter that sources the lensing potential. Second, the DM--DE channel produces a small uniform suppression of the lensing power: at $\xi_{\rm DM} = 1$ (blue), the $\phi\phi$ ratio is $0.9954$ at $\ell=100$ and $0.9980$ at $\ell=1000$, consistent with its $\sigma_8$ suppression. Third, and most strikingly, the DM--$\nu$ channel \emph{enhances} the lensing potential. At the $\sigma_8$-peak coupling $g_{\chi\nu} = 3\times 10^{-4}$ (green), the $\phi\phi$ ratio reaches $1.064$ at $\ell=100$, $1.168$ at $\ell=500$, and $1.129$ at $\ell=1000$. Even at the small-scale-structure bound $g_{\chi\nu} \simeq 10^{-5}$, the enhancement is still $\sim 5\%$ at $\ell=1000$. This is the sharpest CMB-based signature of any channel at currently viable couplings. The physical origin is the same mechanism that produces the $\sigma_8$ peak: the drag damps neutrino free-streaming, so neutrinos cluster with CDM and the total matter perturbation sourcing the lensing potential is larger. At stronger coupling, the drag locks CDM to the free-streaming neutrino velocity, suppressing the total matter perturbation, as seen in the excluded $g_{\chi\nu} = 10^{-2}$ curve (red), which drops to $0.31$ at $\ell = 1000$.

The lensing potential therefore provides the sharpest CMB-based constraint on $g_{\chi\nu}$. The current Planck lensing precision is $\sim 2\%$ per multipole in the range $\ell \in [100,1000]$; this does not reach the $\sim 5\%$ DM--$\nu$ enhancement at $\ell = 1000$ individually, but the signal is coherent across the multipole range and a full likelihood analysis will have sensitivity to it. The projected CMB-S4 lensing sensitivity of $\sim 0.3\%$ over the same range will probe the DM--$\nu$ coupling below the current small-scale structure bound. This makes the lensing potential a complementary constraint on $g_{\chi\nu}$ independent of the $P(k)$ bound, and it makes the DM--$\nu$ channel the only one of the three with a detectable CMB signature at currently allowed parameter values. Crucially, because the $g_{\chi\nu}$ enhancement and the $m_\nu$ suppression act in opposite directions and are parallel in amplitude (as shown in Fig.~\ref{fig:FIG1}), the lensing signature provides a probe of the coupling that is nearly orthogonal to the sum of neutrino masses. Furthermore, the positive sign of the predicted enhancement distinguishes it from the suppression expected from baryonic feedback, making it a robust target for future high-resolution lensing surveys.

\subsection{Growth history and redshift-space distortions}
\label{sec:results-fsigma8}

\begin{figure}[t]
	\centering
	\includegraphics[width=\columnwidth]{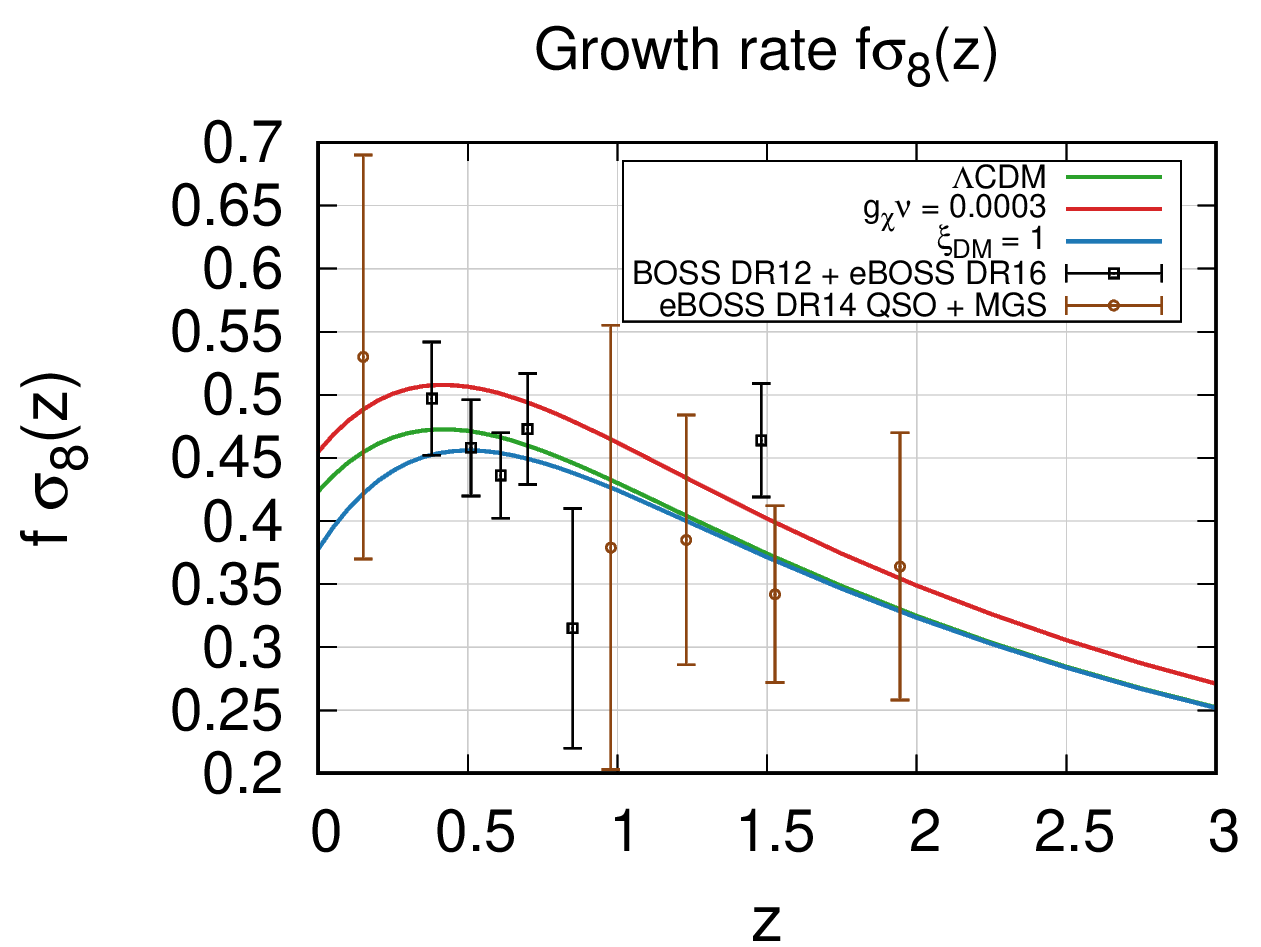}
	\caption{Growth rate $f\sigma_8(z)$ for $\Lambda$CDM (green), the
		$\sigma_8$-enhancing coupling $g_{\chi\nu} = 3\times 10^{-4}$ (red),
		and the saturated DM--DE coupling $\xi_{DM} = 1$ (blue). Data points:
		BOSS DR12~\cite{Alam2017}, eBOSS DR16 LRG~\cite{Bautista2021} and
		ELG~\cite{deMattia2021}, eBOSS DR16 QSO
		consensus~\cite{Hou2021} (black squares); eBOSS DR14 QSO
		tomographic~\cite{Zhao2019} and SDSS
		MGS~\cite{Howlett2015} (brown circles).}
	\label{fig:FIG4}
\end{figure}

Figure~\ref{fig:FIG4} shows the growth rate $f\sigma_8(z)$ for the $\Lambda$CDM baseline, the $\sigma_8$-enhancing coupling $g_{\chi\nu} = 3\times 10^{-4}$, and the saturated DM--DE coupling
$\xi_{DM} = 1$. We compare against the RSD measurements from BOSS DR12, eBOSS DR16 LRG/ELG, and the eBOSS DR16 QSO consensus (black squares), together with the earlier eBOSS DR14 QSO tomographic analysis and the SDSS MGS measurement (brown circles). We emphasize that the \(g_{\chi\nu} = 3\times 10^{-4}\) curve is shown purely to
illustrate the distinct shape of the growth signature in this regime; as shown in Sec.~\ref{sec:results-pk}, this parameter point is already excluded by small-scale structure.

The $g_{\chi\nu} = 3\times 10^{-4}$ curve (red) lies approximately $7\%$ above $\Lambda$CDM at every redshift, reflecting the constant multiplicative enhancement of the growth function that we observed in $\sigma_8$. At the effective redshifts of the BOSS and eBOSS LRG measurements ($z \sim 0.4$--$0.7$), the coupled model sits closer to the central values than $\Lambda$CDM does, though both remain within the $1\sigma$ error bars. At $z = 1.48$, both models lie below the eBOSS QSO measurement, which is a known tension present in $\Lambda$CDM itself. The $\xi_{DM} = 1$ curve (blue) is indistinguishable from $\Lambda$CDM at the precision of the current data.

We do not perform a full covariance analysis in this work. The RSD measurements we compare against are derived from overlapping galaxy samples with correlated cosmological priors, and a proper likelihood would require the full covariance matrix. We note only that the
qualitative trend — mild improvement at intermediate redshifts, unchanged at high redshift — is consistent with the model prediction. Future DESI full-shape RSD measurements will provide the sharper test.

\subsection{Matter power spectrum and small-scale constraint}
\label{sec:results-pk}

\begin{figure}[t]
	\centering
	\includegraphics[width=\columnwidth]{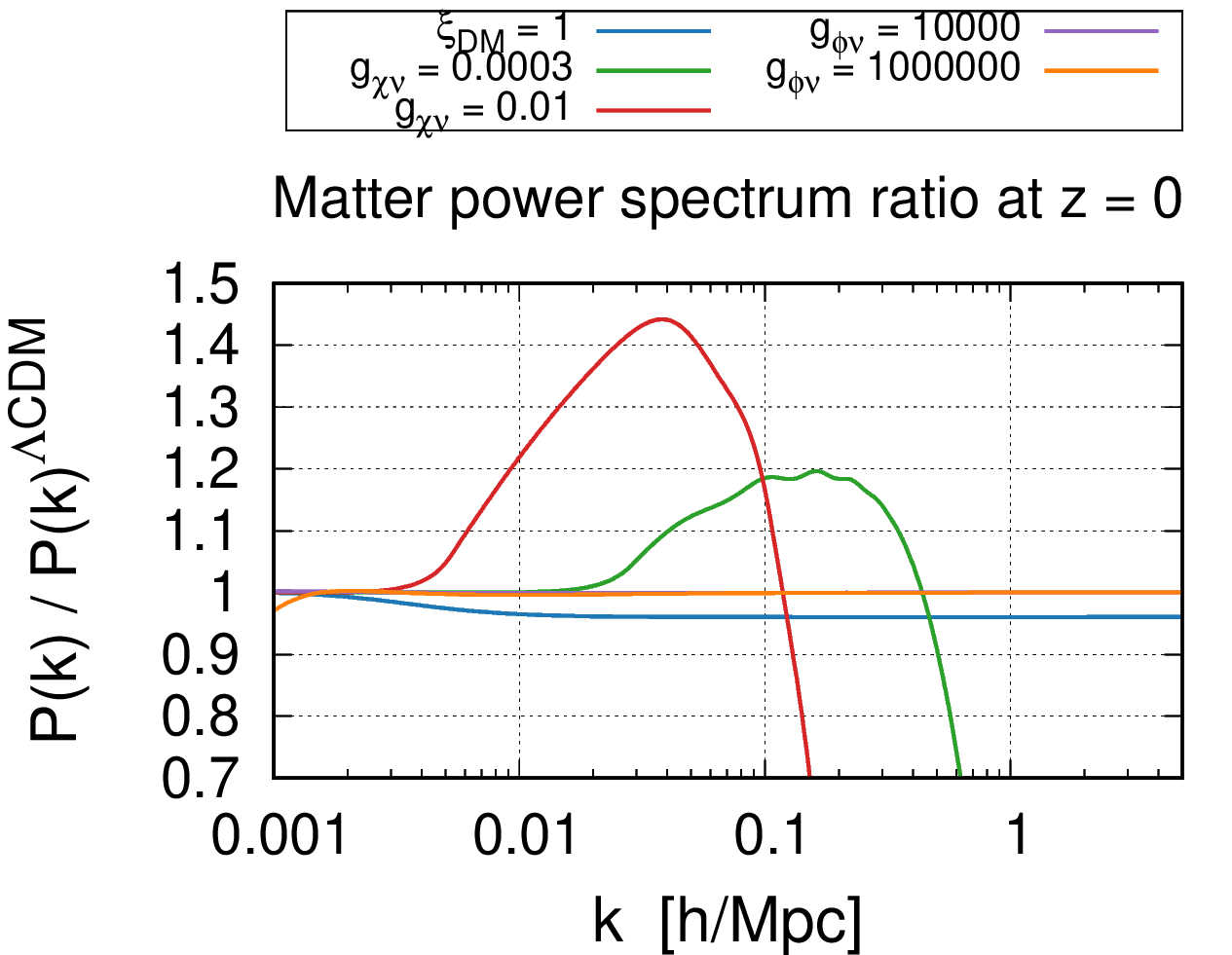}
	\caption{Linear matter power spectrum ratio at $z = 0$. The DM--$\nu$
		coupling produces a strongly scale-dependent modification: enhancement
		at $k \sim 0.1$--$0.3$~$h$/Mpc and rapid suppression at
		$k > 0.5$~$h$/Mpc. The suppression is a generic feature of momentum
		exchange between CDM and a free-streaming species and constrains
		$g_{\chi\nu} \lesssim 10^{-5}$.}
	\label{fig:FIG5}
\end{figure}

Figure~\ref{fig:FIG5} shows the linear matter power spectrum ratio at $z = 0$ for the five representative parameter choices. The DM--$\nu$ coupling produces a strongly scale-dependent modification. At
$g_{\chi\nu} = 3\times 10^{-4}$, the power spectrum is enhanced by $\sim 20\%$ at $k \simeq 0.1$~$h$/Mpc, crosses unity at $k \simeq 0.4$~$h$/Mpc, and is suppressed by more than an order of magnitude at $k > 1$~$h$/Mpc. The situation is even more severe for
the larger, \(S_8\)-resolving coupling $g_{\chi\nu} = 10^{-2}$: the power spectrum exhibits a $\sim 20\%$ enhancement at very large scales ($k \simeq 0.01$~$h$/Mpc), before plunging into a $\sim 40\%$
suppression by $k \simeq 0.025$~$h$/Mpc and becoming wildly oscillatory at smaller scales. This extreme scale-dependent distortion rules out the $g_{\chi\nu} = 10^{-2}$ regime entirely, independently of its effect on the integrated amplitude \(\sigma_8\).

The origin of this behaviour is the tight-coupling dynamics of the drag. The rate at which the DM--$\nu$ channel drives $\theta_{\rm cdm}$ toward $\theta_\nu$ scales as $\Gamma \propto g_{\chi\nu}^2 (\rho_\nu +
P_\nu) H$. In the radiation era, $\rho_\nu \propto (1+z)^4$ and $H \propto (1+z)^2$, so the drag rate grows as $(1+z)^6$. Any nonzero coupling therefore enters the tight-coupling regime at sufficiently high redshift. Once $\theta_{\rm cdm}$ is locked to $\theta_\nu$, and since free-streaming damps the neutrino velocity at small scales, the CDM velocity source in the continuity equation vanishes, and CDM perturbations stop growing. Modes that enter the horizon during the epoch when the drag is active are consequently suppressed relative to
$\Lambda$CDM, with the suppression scale set by the coupling strength.

The sensitivity of the constraint to $g_{\chi\nu}$ is severe. From a controlled comparison at fixed $k$-grid, we find $P(k=2\,h/{\rm Mpc})/P_{\Lambda\rm CDM} = 0.99$ at
$g_{\chi\nu} = 10^{-5}$ and $0.56$ at $g_{\chi\nu} = 3\times 10^{-5}$. Galaxy clustering measurements at $k \sim 1$~$h$/Mpc and Lyman-$\alpha$ forest measurements at $k \sim 5$--$50$~$h$/Mpc constrain the linear power to within a few percent over this range, so the DM--$\nu$
coupling is bounded at $g_{\chi\nu} \lesssim 10^{-5}$ at high confidence. The $\sigma_8$ enhancement at
$g_{\chi\nu} = 3\times 10^{-4}$ is therefore excluded by small-scale structure, independent of the $S_8$ tension.

This bound is a direct consequence of the $(1+z)^6$ scaling of the drag rate, and it applies to any momentum-exchange model in which CDM is coupled to a free-streaming species through a rate that grows
faster than the Hubble rate at early times. The specific coefficient of the bound depends on the coupling structure: the enthalpy weighting $\rho_\nu + P_\nu$ that we have derived enhances the drag rate by $4/3$ in the relativistic era relative to a phenomenological
parameterization with $\Gamma \propto \rho_\nu$, tightening the $1\sigma$ bound on $g_{\chi\nu}$ by approximately $15\%$. The qualitative behaviour, however, is set by the cosmological scaling rather than by the microphysical normalization. This is also the reason our framework cannot reach the interaction strengths $u_{\nu\rm DM} \sim 10^{-4}$ that phenomenological DM--$\nu$ elastic scattering scenarios invoke to alleviate the $S_8$ tension~\cite{Zu2026}. A constant elastic scattering cross-section gives $\Gamma \propto n_\nu\sigma \propto (1+z)^3$, whereas the derivative EFT coupling used here scales as $\Gamma \propto (\rho_\nu + P_\nu) H \propto (1+z)^6$ in the radiation era, an enhancement of three additional powers of $(1+z)$. The UV completion and technical naturalness of the coupling therefore come at the price of a much stronger high-redshift constraint, and it is this scaling---rather than any tuning of the microphysics---that rules out the $S_8$-resolving regime in our framework.

The DM--DE and $\nu$--DE channels do not exhibit this small-scale suppression. The DM--DE channel couples to DE, which does not free-stream at the scales relevant for structure formation, and its $P(k)$ modification is a smooth $\sim 2\%$ suppression across all scales. The $\nu$--DE channel modifies the neutrino velocity but
leaves CDM untouched, and its $P(k)$ imprint is below $0.1\%$ for any viable coupling.

\subsection{Enthalpy weighting and the origin of the drag}
\label{sec:results-enthalpy}

\begin{figure}[t]
	\centering
	\includegraphics[width=\columnwidth]{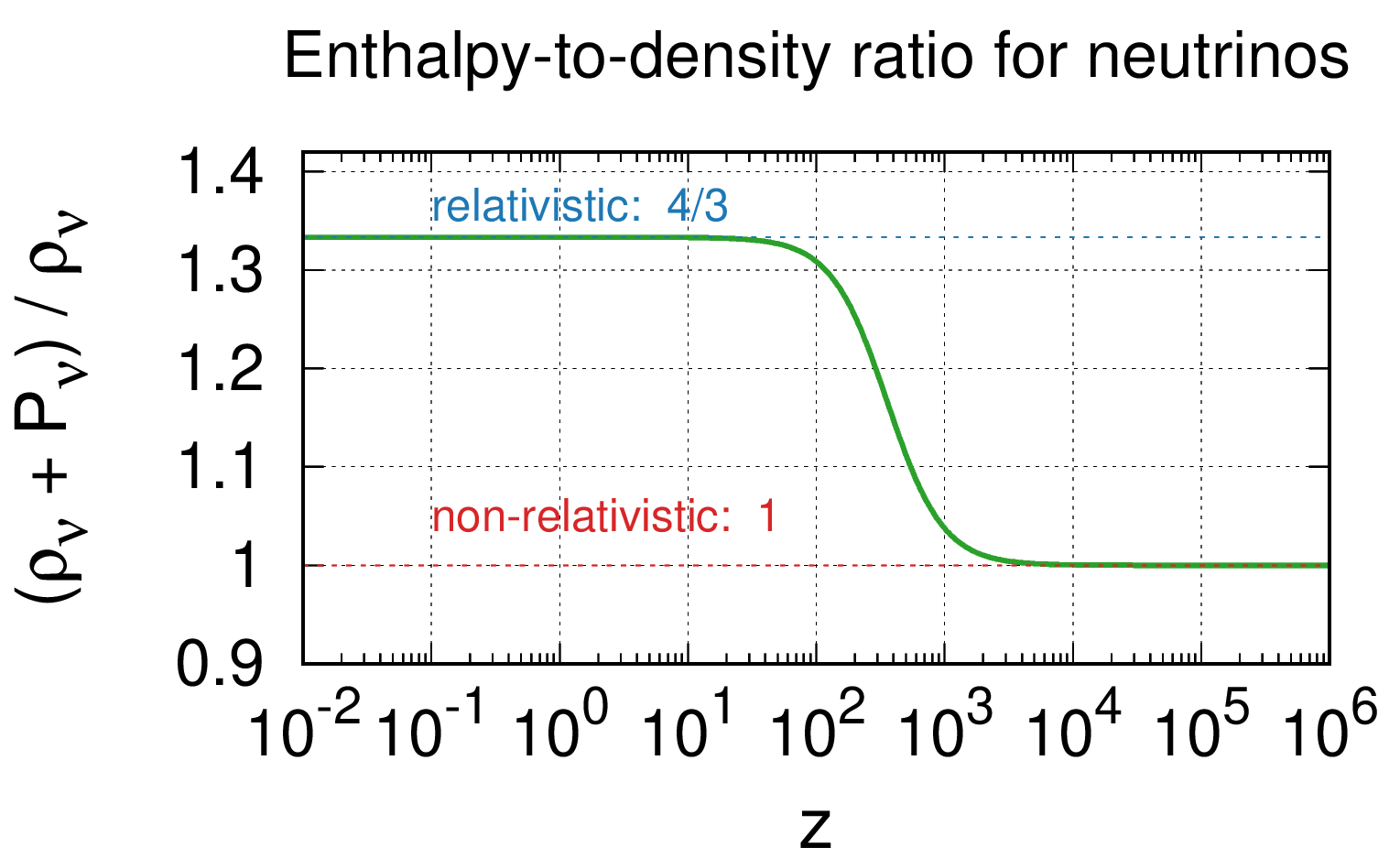}
	\caption{Neutrino enthalpy-to-density ratio $(\rho_\nu +
		P_\nu)/\rho_\nu$ as a function of redshift. The ratio asymptotes to
		$4/3$ in the relativistic era and to $1$ in the non-relativistic era.
		The $4/3$ factor is a direct consequence of the covariant derivation
		of the drag rate (Appendix~\ref{app:derivation}) and distinguishes the
		unified construction from phenomenological parameterizations with
		$\Gamma \propto \rho_\nu$.}
	\label{fig:FIG6}
\end{figure}

Figure~\ref{fig:FIG6} shows the neutrino enthalpy-to-density ratio $(\rho_\nu + P_\nu)/\rho_\nu$ as a function of redshift for the fiducial neutrino mass $m_\nu = 0.06$~eV. The ratio asymptotes to $4/3$ in the
relativistic era ($z \gg 350$) and to $1$ in the non-relativistic era. The transition is smooth and corresponds to the epoch at which the neutrino thermal energy drops below the rest mass.

The $4/3$ factor is the direct and distinctive consequence of the covariant derivation of the drag rate presented in Appendix~\ref{app:derivation}. The drag force on the neutrino fluid is proportional to the enthalpy density $\rho_\nu + P_\nu$, not to the
energy density $\rho_\nu$ alone. In the relativistic limit
$P_\nu = \rho_\nu/3$, so $\rho_\nu + P_\nu = (4/3)\rho_\nu$. This differs from phenomenological interacting-DE parameterizations in
which the drag rate is taken to be proportional to the energy density \(\rho_\nu\) (as in the Dark Scattering models~\cite{Simpson2010} and their CLASS implementations~\cite{Pourtsidou2016}) or treated as a free function of the scale factor~\cite{Linton2018, Linton2022}.

The practical consequence of the $4/3$ enhancement is twofold. First, it strengthens the constraint on $g_{\chi\nu}$ by roughly $15\%$ relative to a rate $\propto \rho_\nu$, as noted in Sec.~\ref{sec:results-pk}. Second, and more importantly, it makes the drag active over a wider redshift range than a naive estimate would suggest. The $4/3$ factor shifts the onset of the tight-coupling regime to slightly higher redshift and therefore increases the amount of growth suppression that the model predicts at small scales. The
enthalpy weighting is thus not a phenomenological detail but a structural feature of the covariant construction, and it enters the observational predictions at the level of precision that is relevant for current data.

For completeness we note that the analogous ratio for CDM is exactly $1$ at all redshifts, since CDM is pressureless. The $4/3$ enhancement is therefore a property of the relativistic neutrino sector specifically, and it does not appear in the DM--DE channel.

\subsection{Summary of observational signatures}
\label{sec:results-summary}

Table~\ref{tab:signatures} summarizes the qualitative behaviour of each channel across the observables we have considered. Three points stand out.

First, the DM--$\nu$ channel is the only one that produces a non-monotonic $\sigma_8$ response, and its $S_8$-enhancing regime is excluded by small-scale structure. The channel is therefore observationally viable only for $g_{\chi\nu} \lesssim 10^{-5}$, where
its imprint on all observables is small.

Second, the $\nu$--DE channel is invisible in the primary CMB, in $\sigma_8$, and in the CMB lensing potential. Its imprint on all current observables is null, as expected from the fact that it does not couple to cold dark matter.

Third, the DM--DE channel is the only channel that produces a non-zero \(\sigma_8\) shift at currently allowed couplings, but its effect is bounded at the $2\%$ level, which is insufficient to resolve the \(S_8\) tension. This reproduces the earlier result in the unified framework~\cite{Farhan:2026sym}.

The combined picture is that the unified dark sector we have constructed does not resolve the $S_8$ tension. The couplings that would be needed to produce the required suppression are excluded by small-scale structure, and the viable parameter space produces $\sigma_8$ shifts at the level of a few percent or less. The paper's
contribution is the identification of the specific mechanism — the $(1+z)^6$ scaling of the drag rate combined with the free-streaming
suppression of the CDM velocity — that rules out the interesting parameter regime, together with the derivation of the $4/3$ enthalpy enhancement that distinguishes this construction from phenomenological
alternatives.

\begin{table}[t]
	\centering
	\caption{Qualitative observational signature of each momentum-exchange channel. Entries indicate the sign and approximate magnitude of the effect at the given precision. A dash (---) indicates no visible effect
	at currently viable couplings.}
	\label{tab:signatures}
	\begin{ruledtabular}
		\begin{tabular}{l c c p{1.3cm} p{1.3cm}}
			Channel & $\sigma_8$ & CMB & Lensing & $P(k)$ \\
			\hline
			DM--DE ($\xi_{DM}$)     & $-2\%$ (sat.) & ---  & weak    & $-2\%$ (flat) \\
			$\nu$--DE ($g_{\phi\nu}$) & $\ll 0.1\%$   & ---  & --- & --- \\
			DM--$\nu$ ($g_{\chi\nu}$) & non-monotonic & ---  & strong enh.    & strong supp. \\
		\end{tabular}
	\end{ruledtabular}
\end{table}
	
\section{Conclusions}
\label{sec:conclusions}

We have extended the $Z_4 \times Z_2^{\rm dark}$-symmetric inert doublet plus singlet framework \cite{Naeem:2026layered,Farhan:2026sym} to include two derivative couplings between the neutrino energy current and, respectively, the dark energy pNGB and the scalar dark matter field. Together with the DM--DE operator of the original model, this yields a unified dark sector in which dark matter, dark energy, and neutrinos exchange momentum through three shift-symmetric dimension-six channels. All three interactions are protected by the same symmetries that stabilize the individual dark sector masses, and none of them generates large radiative corrections to $m_\phi$, $m_\chi$, or the neutrino masses.

The central theoretical result of this work is the covariant structure of the neutrino momentum-transfer force. Insisting that the transport rate $\Gamma_{\nu X}$ be a genuine rate --- independent of the fluid state, and not a normalization-dependent combination of fluid variables --- forces the four-force density to be proportional to the neutrino enthalpy density $\rho_\nu + P_\nu$. This produces a $4/3$ enhancement of the effective drag in the relativistic era and cannot be reproduced by any phenomenological parameterization in which the rate is taken to be proportional to $\rho_\nu$ alone, or treated as an arbitrary function of the scale factor. The enhancement follows directly from the covariant conservation equation $\nabla_\mu T^{\mu\nu} = Q^\nu$ with $u_\mu Q^\mu = 0$, and is the neutrino-sector analogue of the fact that the momentum density of a relativistic fluid is $(\rho+P)u^\mu$ rather than $\rho u^\mu$. The associated enthalpy-to-density ratio $(\rho_\nu+P_\nu)/\rho_\nu$, which interpolates from $4/3$ in the relativistic era to $1$ in the non-relativistic era, tightens the bound on the DM--$\nu$ coupling by approximately $15\%$ relative to a rate proportional to $\rho_\nu$ alone, and shifts the onset of the tight-coupling regime to slightly higher redshift.

The three channels leave qualitatively distinct signatures on cosmological observables. The DM--DE drag slows the dominant clustering component directly and produces a monotonic suppression of $\sigma_8$ that saturates at approximately $-2\%$ in the strong-coupling limit, consistent with the earlier finding that this channel alone cannot resolve the $S_8$ tension. The DM--$\nu$ channel is non-monotonic: at small coupling the damping of neutrino free-streaming enhances total matter clustering by up to $7.4\%$ relative to $\Lambda$CDM, with the peak at $g_{\chi\nu} \simeq 3\times 10^{-4}$ and $\sigma_8 = 0.871$ for the fiducial $m_\nu = 0.06$~eV. To instead suppress \(\sigma_8\) into the weak-lensing band, a much larger coupling \(g_{\chi\nu} \sim 10^{-3}\) is required. However, the \((1+z)^6\) scaling of the drag rate in the radiation era drives the DM--$\nu$ system into the tight-coupling regime at sufficiently high redshift for any nonzero coupling. The resulting suppression of the linear power spectrum at $k \gtrsim 1\,h/{\rm Mpc}$ constrains $g_{\chi\nu} \lesssim 10^{-5}$ at high confidence, excluding both the enhancing peak and the \(S_8\)-resolving suppressing regime. At the viable couplings the DM--$\nu$ channel therefore shifts $\sigma_8$ by at most a fraction of a percent. The $\nu$--DE channel leaves the primary CMB, $\sigma_8$, and the CMB lensing potential essentially unchanged even at saturation, because it couples two subdominant species without touching the cold dark matter that sources the Weyl potential. Its cosmological signature is therefore not in the CMB or in the clustering amplitude. The DM--$\nu$ channel, by contrast, leaves a distinctive $\phi\phi$ enhancement that survives at currently viable couplings: $\sim 5\%$ at $\ell = 1000$ for the small-scale-structure bound $g_{\chi\nu} \simeq 10^{-5}$, growing to $\sim 17\%$ at $\ell = 500$ for the $\sigma_8$-peak coupling $g_{\chi\nu} \simeq 3 \times 10^{-4}$. This enhancement is a target for CMB-S4, which will probe the DM--$\nu$ coupling independently of the $P(k)$ bound, and it makes the lensing potential the only observable in which the unified dark sector leaves a channel-specific CMB signature at viable parameter values.

The combined picture is that the unified dark sector we have constructed does not resolve the $S_8$ tension. The couplings that would be needed to produce the required suppression of the clustering amplitude are excluded by the small-scale structure of the matter power spectrum, independently of the weak-lensing measurements that motivate the tension in the first place. The viable parameter space of the model produces $\sigma_8$ shifts at the level of a few percent or less, dominated by the saturated DM--DE channel. This mirrors the conclusion of the original $Z_4$-IDSM construction and extends it: the addition of the two neutrino channels does not open a new route to resolving the tension, because the only mechanism that enhances $\sigma_8$ at small coupling --- the damping of neutrino free-streaming by a moderate DM--$\nu$ drag --- is precisely the mechanism that is ruled out by the same small-scale data. 

This result sharply distinguishes our UV-complete EFT framework from phenomenological DM--$\nu$ elastic scattering models~\cite{Zu2026}, which can alleviate the $S_8$ tension using a cross-section scaling as \((1+z)^3\). By contrast, the derivative dimension-six operator required by our symmetries scales as \((1+z)^6\), forcing the dark sectors into tight coupling at high redshifts and making the $S_8$-resolving regime strictly inaccessible. The technical naturalness and UV completion of the coupling are therefore fundamentally at odds with using it to solve the $S_8$ tension.

The results also identify two features of the unified framework that are worth pursuing independently of the $S_8$ question. The first is the degeneracy structure between the DM--$\nu$ coupling and the total neutrino mass. The five $m_\nu$ curves in Fig.~\ref{fig:FIG1} are parallel across the entire coupling range, with vertical offsets that match the $\Lambda$CDM neutrino-mass sensitivity to within a few percent, and with a peak position that is unchanged as $m_\nu$ varies. This means that the coupling and the neutrino mass modify $\sigma_8$ independently and can in principle be constrained separately by the same data; a full MCMC analysis combining Planck, DESI DR2 BAO, and KiDS-1000 cosmic shear would determine whether this separability survives the degeneracies of the full cosmological parameter space. The second is the CMB lensing signature of the DM--$\nu$ channel. A momentum-exchange coupling that leaves the primary CMB and the linear matter power spectrum largely untouched but enhances the lensing potential by roughly five percent at currently allowed parameter values is a rare configuration, and it provides a clean, channel-specific target for the next generation of CMB experiments.

Several extensions of this work are natural. The most immediate is the full likelihood analysis just mentioned; with the low-$\ell$ and high-$\ell$ Planck likelihoods, BAO measurements from DESI DR2, and cosmic shear from KiDS-1000, one can test whether the three couplings are constrained separately or only in combination, and what independent bound the neutrino-specific observables place on $g_{\phi\nu}$. A Bayesian comparison against the phenomenological DM--$\nu$ elastic scattering scenarios ~\cite{Zu2026} would additionally quantify the statistical preference between a UV-motivated derivative coupling (which predicts a CMB lensing enhancement but cannot resolve \(S_8\)) and a phenomenological cross-section (which can resolve \(S_8\) but requires a highly suppressed or redshift-dependent cross-section to evade small-scale bounds). A second direction is the nonlinear regime: the scale-dependent enhancement of the linear power spectrum at $g_{\chi\nu} \sim 10^{-4}$---where the enhancement at $k \sim 0.1$--$0.3\,h/{\rm Mpc}$ reaches $20\%$---will feed back into the one-loop and halo-model corrections, and the resulting shape change in $P(k)$ at intermediate scales may be a sharper discriminator of the model than the linear $\sigma_8$ alone, even at the couplings that are currently allowed. A third direction, more model-building than phenomenology, is to examine whether the $\nu$--DE coupling can be realized with a natural UV completion in which the $U(1)_S$ breaking scale $v_s$ is not of order $M_{\rm Pl}$, thereby relaxing the inflationary dilution assumption made in Sec.~\ref{sec:potential}. We leave these questions to future work.

\appendix

\section{Derivation of the Scotogenic Neutrino Mass}
\label{app:scotogenic}

We provide a detailed derivation of the one-loop neutrino mass matrix in the scotogenic model, starting from the Lagrangian and performing the loop integral explicitly.

\subsection{The Lagrangian and Feynman rules}

The relevant terms in the neutrino sector are
\begin{equation}
	\mathcal{L}_{\nu}
	=
	y_{ik} \bar{L}_{L_i} \tilde{H}_2 N_{R_k}
	+
	\frac{1}{2} M_{N_k} \bar{N}_{R_k}^c N_{R_k}
	+
	\text{h.c.},
	\label{eq:app_neutrino_lagrangian}
\end{equation}
where $L_{L_i}$ are the left-handed lepton doublets, $\tilde{H}_2 = i\sigma_2 H_2^*$ is the charge-conjugated inert doublet, and $N_{R_k}$ are the right-handed neutrino singlets. The neutral components of the inert doublet are denoted $H_2^0$ (CP-even) and $A^0$ (CP-odd). The mass splitting between these states is
\begin{equation}
	m_{A^0}^2 - m_{H_2^0}^2 = \lambda_5 v^2,
	\label{eq:app_mass_splitting}
\end{equation}
where $v = 174\,\text{GeV}$ is the electroweak VEV.

\subsection{The one-loop amplitude}

The amplitude for the neutrino self-energy is
\begin{equation}
	-i \Sigma_{ij}(p) = \int \frac{d^4 q}{(2\pi)^4} \frac{(-i y_{ik})(-i y_{jk})}{(q^2 - m_{H_2^0}^2)} \frac{i M_{N_k}}{q^2 - M_{N_k}^2} - (H_2^0 \to A^0),
	\label{eq:app_loop_amplitude}
\end{equation}
where $p$ is the external momentum, $q$ is the loop momentum, and we have suppressed the spinor structure for brevity. The fermion propagator for the Majorana neutrino is
\begin{equation}
	S_F(q) = \frac{i(\not{q} + M_{N_k})}{q^2 - M_{N_k}^2},
\end{equation}
and the mass insertion on the internal fermion line gives the factor $M_{N_k}$. Equation~\eqref{eq:app_loop_amplitude} is a difference of two bubble diagrams, each containing one scalar and one fermion propagator. The two contributions cancel in the limit $\lambda_5 \to 0$, as required for a radiatively generated mass.

\subsection{Wick rotation}

To evaluate each loop integral, we perform a Wick rotation to Euclidean space. The integration contour in the complex $q^0$ plane is rotated: $q^0 \to i q_E^0$, so that $q^2 = (q^0)^2 - \mathbf{q}^2 \to -q_E^2$, where $q_E^2 = (q_E^0)^2 + \mathbf{q}^2$. The measure transforms as
\begin{equation}
	d^4 q = i \, d^4 q_E, \qquad \int \frac{d^4 q}{(2\pi)^4} = i \int \frac{d^4 q_E}{(2\pi)^4}.
\end{equation}
Applying this to Eq.~\eqref{eq:app_loop_amplitude} gives
\begin{equation}
	\begin{split}
		-i \Sigma_{ij}(p)
		&=
		i \int \frac{d^4 q_E}{(2\pi)^4}
		\Bigg[
		\frac{(-i y_{ik})(-i y_{jk})}{q_E^2 + m_{H_2^0}^2}
		\frac{i M_{N_k}}{q_E^2 + M_{N_k}^2}
		\\
		&\qquad
		-
		(H_2^0 \to A^0)
		\Bigg].
	\end{split}
\end{equation}
Simplifying the factors of $i$:
\begin{equation}
	\begin{split}
		\Sigma_{ij}(p)
		&=
		- y_{ik} y_{jk} M_{N_k} \int \frac{d^4 q_E}{(2\pi)^4}
		\Bigg[
		\frac{1}{(q_E^2 + m_{H_2^0}^2)(q_E^2 + M_{N_k}^2)}
		\\
		&\qquad
		-
		\frac{1}{(q_E^2 + m_{A^0}^2)(q_E^2 + M_{N_k}^2)}
		\Bigg].
	\end{split}
	\label{eq:app_wick_rotated}
\end{equation}

\subsection{The loop integral}

Each term in Eq.~\eqref{eq:app_wick_rotated} is a two-propagator Euclidean integral of the form
\begin{equation}
	J(m_1^2, m_2^2) = \int \frac{d^4 q_E}{(2\pi)^4} \frac{1}{(q_E^2 + m_1^2)(q_E^2 + m_2^2)}.
	\label{eq:app_loop_integral}
\end{equation}
Using Feynman parameters,
\begin{equation}
	\frac{1}{(q_E^2 + m_1^2)(q_E^2 + m_2^2)} = \int_0^1 dx \frac{1}{[q_E^2 + x m_1^2 + (1-x) m_2^2]^2},
\end{equation}
and
\begin{equation}
	\int \frac{d^4 q_E}{(2\pi)^4} \frac{1}{(q_E^2 + \Delta)^2} = \frac{1}{16\pi^2} \frac{1}{\Delta}.
\end{equation}
Performing the $x$-integral gives
\begin{equation}
	J(m_1^2, m_2^2) = \frac{1}{16\pi^2} \frac{\ln(m_1^2/m_2^2)}{m_1^2 - m_2^2}.
	\label{eq:app_loop_result}
\end{equation}
This is dimensionally correct ($J$ has mass dimension $[\text{mass}]^{-2}$) and symmetric under interchange of $m_1^2 \leftrightarrow m_2^2$.

\subsection{The final mass matrix}

Substituting Eq.~\eqref{eq:app_loop_result} into Eq.~\eqref{eq:app_wick_rotated} with $m_{H_2^0}$ and $m_{A^0}$ for the scalar masses and $M_{N_k}$ for the fermion, we obtain
\begin{equation}
	\Sigma_{ij}(p) = - y_{ik} y_{jk} M_{N_k} \left[ J(m_{H_2^0}^2, M_{N_k}^2) - J(m_{A^0}^2, M_{N_k}^2) \right].
\end{equation}
Using the mass splitting relation Eq.~\eqref{eq:app_mass_splitting} and writing each term in the form
\begin{equation}
	J(m^2, M^2) = \frac{1}{16\pi^2} \frac{\ln(m^2/M^2)}{m^2 - M^2},
\end{equation}
the light neutrino mass matrix becomes
\begin{equation}
	\begin{split}
		m_{\nu}^{ij}
		&=
		\sum_k \frac{y_{ik} y_{jk} M_{N_k}}{16\pi^2}
		\Bigg[
		\frac{m_{H_2^0}^2}{m_{H_2^0}^2 - M_{N_k}^2}
		\ln\left(\frac{m_{H_2^0}^2}{M_{N_k}^2}\right) \\
		&\quad
		- \frac{m_{A^0}^2}{m_{A^0}^2 - M_{N_k}^2}
		\ln\left(\frac{m_{A^0}^2}{M_{N_k}^2}\right)
		\Bigg],
	\end{split}
	\label{eq:app_final_mass}
\end{equation}
where the sign has been chosen so that each term is manifestly finite and the result reduces to the standard form below. In the limit $m_{A^0}^2, m_{H_2^0}^2 \ll M_{N_k}^2$, Eq.~\eqref{eq:app_final_mass} reduces to
\begin{equation}
	m_\nu^{ij} \simeq \frac{\lambda_5 v^2}{16\pi^2} \sum_k \frac{y_{ik} y_{jk}}{M_{N_k}} \left[ \ln\left(\frac{M_{N_k}^2}{m_0^2}\right) - 1 \right],
\end{equation}
where $m_0^2 = (m_{H_2^0}^2 + m_{A^0}^2)/2$, consistent with literature~\cite{Ma2006}.

A parameter scan of the scotogenic model establishes the relation
\begin{equation}
	\lambda_5 = 3.1 \times 10^{-9} \left( \frac{m_{\nu_e}}{\text{eV}} \right),
	\label{eq:app_lambda5_relation}
\end{equation}
which we use to constrain the viable parameter space of our model.

\section{Covariant derivation of the neutrino momentum-transfer structure}
\label{app:derivation}

In this appendix, $T^{\mu\nu}$, $u^\mu$, and $Q^\mu$ denote the neutrino stress-energy tensor, four-velocity, and four-force density, respectively. The neutrino label is suppressed for clarity; it should not be confused with a spacetime index.

We formulate the neutrino sector as a perfect fluid with stress-energy tensor
\begin{equation}
	T^{\mu\nu}
	=
	(\rho_\nu + P_\nu) u^\mu u^\nu
	+
	P_\nu g^{\mu\nu},
	\label{eq:nu_stress_tensor}
\end{equation}
where $u^\mu u_\mu = -1$. The orthogonal projector onto the spatial hypersurface in the neutrino rest frame is
\begin{equation}
	\Delta^{\mu\nu}
	=
	g^{\mu\nu} + u^\mu u^\nu.
	\label{eq:nu_projector}
\end{equation}

The covariant conservation equation in the presence of an interaction is
\begin{equation}
	\nabla_\mu T^{\mu\nu}
	=
	Q^\nu,
	\label{eq:nu_conservation}
\end{equation}
where $Q^\nu$ is the four-force density acting on the neutrino fluid. We impose pure momentum exchange by requiring
\begin{equation}
	u_\mu Q^\mu = 0.
	\label{eq:pure_momentum}
\end{equation}
This condition guarantees that the interaction transfers momentum but no energy in the local neutrino rest frame and therefore does not modify the homogeneous background energy-density evolution.

For an isotropic drag interaction between neutrinos and another component $X$, covariance and linearity in the relative velocity imply the leading-order four-force
\begin{equation}
	Q^\mu
	=
	-(\rho_\nu + P_\nu)\Gamma_{\nu X}
	\Delta^\mu{}_\alpha
	\left(
	u^\alpha - u_X^\alpha
	\right),
	\label{eq:Q_general}
\end{equation}
where $\Gamma_{\nu X}$ is the momentum-transfer rate in the neutrino rest frame. The corresponding force on the second component follows from total energy-momentum conservation,
\begin{equation}
	Q_X^\mu = -Q^\mu.
	\label{eq:momentum_conservation}
\end{equation}

The enthalpy prefactor $\rho_\nu + P_\nu$ in Eq.~\eqref{eq:Q_general} is not a normalization choice. Insisting that the transport coefficient $\Gamma_{\nu X}$ be a genuine rate --- with units of inverse time, independent of the fluid state --- forces the four-force density to be proportional to the momentum density of the neutrino fluid, $(\rho_\nu+P_\nu)u^\mu$, and not to the energy density $\rho_\nu u^\mu$ alone. Absorbing the prefactor into a redefined rate would make that rate a function of the fluid state, which is not what a microscopic transport coefficient should be. The spatial projection of the conservation equation gives the covariant Euler equation
\begin{equation}
	(\rho_\nu + P_\nu)a^\mu
	+
	\Delta^{\mu\nu}\nabla_\nu P_\nu
	=
	\Delta^{\mu}{}_\alpha Q^\alpha,
	\label{eq:nu_euler_covariant}
\end{equation}
where
\begin{equation}
	a^\mu
	\equiv
	u^\alpha\nabla_\alpha u^\mu
\end{equation}
is the neutrino four-acceleration. Substitution of Eq.~\eqref{eq:Q_general} therefore gives
\begin{equation}
	(\rho_\nu + P_\nu)a^\mu
	+
	\Delta^{\mu\nu}\nabla_\nu P_\nu
	=
	-(\rho_\nu + P_\nu)\Gamma_{\nu X}
	\Delta^\mu{}_\alpha
	(u^\alpha - u_X^\alpha).
	\label{eq:nu_euler_drag_covariant}
\end{equation}

Dividing by the neutrino enthalpy density yields
\begin{equation}
	a^\mu
	+
	\frac{\Delta^{\mu\nu}\nabla_\nu P_\nu}
	{\rho_\nu + P_\nu}
	=
	-\Gamma_{\nu X}
	\Delta^\mu{}_\alpha
	(u^\alpha - u_X^\alpha).
	\label{eq:nu_euler_final_covariant}
\end{equation}

Thus the interaction acts purely as a drag on the relative velocity. No energy-transfer term is generated at the homogeneous level because of Eq.~\eqref{eq:pure_momentum}.

\subsection{Linear perturbations}

In Newtonian gauge,
\begin{equation}
	ds^2
	=
	a^2(\tau)
	\left[
	-(1+2\Psi)d\tau^2
	+
	(1-2\Phi)\delta_{ij}dx^i dx^j
	\right].
\end{equation}
At linear order, the spatial part of the relative four-velocity is proportional to the difference of the velocity divergences,
\begin{equation}
	\Delta_{i\alpha}
	(u^\alpha - u_X^\alpha)
	\longrightarrow
	-\frac{i k_i}{a k^2}
	(\theta_\nu - \theta_X).
\end{equation}

The spatial projection of Eq.~\eqref{eq:nu_euler_final_covariant} therefore gives the Euler equation
\begin{equation}
	\theta_\nu'
	+
	\mathcal H(1-3w_\nu)\theta_\nu
	-
	\frac{c_{s,\nu}^2}{1 + w_\nu}k^2\delta_\nu
	+
	k^2\Psi
	=
	-a\Gamma_{\nu X}
	(\theta_\nu-\theta_X).
	\label{eq:nu_euler_drag}
\end{equation}

The corresponding equation for the interacting component $X$ follows from Eq.~\eqref{eq:momentum_conservation}. For example, when $X$ is dark energy,
\begin{equation}
	\begin{split}
		\theta_{\rm DE}'
		+ &\quad
		\mathcal H(1-3c_s^2)\theta_{\rm DE}
		-
		\frac{c_s^2}{1+w_{\rm DE}}
		k^2\delta_{\rm DE}
		+
		k^2\Psi
		\\
		&\quad
		=
		-a
		\frac{\rho_\nu + P_\nu}
		{\rho_{\rm DE}(1+w_{\rm DE})}
		\Gamma_{\nu{\rm DE}}
		(\theta_{\rm DE}-\theta_\nu).
	\end{split}
	\label{eq:de_backreaction}
\end{equation}

Likewise, for dark matter,
\begin{equation}
	\theta_{\rm DM}'
	+
	\mathcal H\theta_{\rm DM}
	+
	k^2\Psi
	=
	a
	\Gamma_{\rm DM-\nu}
	(\theta_\nu-\theta_{\rm DM}).
\end{equation}

\subsection{Effective transport rate and cosmological closure}
\label{sec:transport}

The derivative interaction between the pNGB and the neutrino enthalpy current may be written schematically as
\begin{equation}
	\mathcal L_{\nu-\phi}
	=
	\frac{g_{\phi\nu}}{\Lambda^2}
	\partial_\mu\phi
	\delta J^\mu,
	\quad
	\delta J^\mu
	\equiv
	J^\mu - \bar J^\mu,
	\quad
	J^\mu
	\equiv
	(\rho_\nu + P_\nu)u^\mu ,
	\label{eq:nu_enthalpy_interaction}
\end{equation}
where the background subtraction $\delta J^\mu = J^\mu - \bar J^\mu$ is the renormalization condition introduced in Sec.~II.C.2. The operator is manifestly invariant under the pNGB shift symmetry $\phi \to \phi + c$.

At the level of the covariant fluid equations, the interaction generates a force orthogonal to the neutrino four-velocity. Writing
\begin{equation}
	\Delta^\mu{}_\alpha
	=
	\delta^\mu{}_\alpha
	+
	u^\mu u_\alpha,
\end{equation}
the force takes the form
\begin{equation}
	Q^\mu
	=
	-\frac{g_{\phi\nu}}{\Lambda^2}
	(\rho_\nu+P_\nu)
	\Delta^\mu{}_\alpha \partial^\alpha\phi ,
	\label{eq:force_from_operator}
\end{equation}
up to the details of the underlying fluid action and the normalization convention for the interaction current.

Variation with respect to the pNGB gives
\begin{equation}
	\Box\phi
	=
	-\frac{g_{\phi\nu}}{\Lambda^2}
	\nabla_\mu \delta J^\mu .
	\label{eq:scalar_from_operator}
\end{equation}
For a perfect fluid,
\begin{equation}
	\nabla_\mu \delta J^\mu
	=
	u^\mu\nabla_\mu\delta(\rho_\nu+P_\nu)
	+
	(\rho_\nu+P_\nu)\nabla_\mu \delta u^\mu ,
\end{equation}
where the background contribution vanishes because $\delta J^\mu = 0$ on FRW. In the subhorizon limit the velocity-divergence contribution provides the leading perturbative source under the usual quasi-static assumptions. Importantly, this step determines the response of the scalar field to the neutrino perturbations, but does not by itself fix a local two-fluid drag coefficient.

We therefore parameterize the momentum transfer in the most general linear, covariant pure-drag form,
\begin{equation}
	Q^\mu
	=
	-(\rho_\nu+P_\nu)
	\Gamma_{\nu X}
	\Delta^\mu{}_\alpha
	(u^\alpha-u_X^\alpha),
	\label{eq:Gamma_definition}
\end{equation}
where $X$ denotes the fluid exchanging momentum with the neutrinos. This form follows from covariance, linearity in the relative velocity, and the requirement $u_\mu Q^\mu=0$.

The microscopic operator determines the dimensional dependence of the corresponding transport coefficient. For the dimension-six interaction in Eq.~\eqref{eq:nu_enthalpy_interaction}, the leading EFT scaling may be written as
\begin{equation}
	\Gamma_{\nu X}
	\sim
	C_{\nu X}
	\frac{g_{\nu X}^2}{\Lambda^4}
	(\rho_\nu+P_\nu)
	\mathcal E_X ,
	\label{eq:general_rate}
\end{equation}
where $C_{\nu X}$ is a dimensionless transport coefficient and $\mathcal E_X$ denotes the characteristic inverse timescale associated with the macroscopic interaction process. The factor $g_{\nu X}^2/\Lambda^4$ follows from the dimension-six normalization, while the remaining dependence specifies the phenomenological transport closure adopted below.

For the cosmological implementation considered here, we adopt an adiabatic closure in which the characteristic macroscopic timescale is set by the Hubble expansion,
\begin{equation}
	\mathcal E_X = H(a).
	\label{eq:adiabatic_scale}
\end{equation}
The resulting effective rates are therefore
\begin{equation}
	\Gamma_{\nu-\rm DE}(a)
	=
	C_{\phi\nu}
	\frac{g_{\phi\nu}^2}{\Lambda^4}
	(\rho_\nu + P_\nu)H(a),
	\label{eq:Gamma_nu_DE_final}
\end{equation}
and
\begin{equation}
	\Gamma_{\rm DM-\nu}(a)
	=
	C_{\chi\nu}
	\frac{g_{\chi\nu}^2}{\Lambda^4}
	(\rho_\nu + P_\nu)H(a).
	\label{eq:Gamma_DM_nu_final}
\end{equation}

The coefficients $C_{\phi\nu}$ and $C_{\chi\nu}$ absorb phase-space factors, numerical normalization conventions, and other short-distance transport details, and are absorbed into the effective couplings used in the numerical analysis.

A robust consequence of the covariant fluid description is the enthalpy weighting of the momentum-transfer force,
\begin{equation}
	Q^\mu
	\propto
	(\rho_\nu+P_\nu)
	\Delta^\mu{}_\alpha
	(u^\alpha-u_X^\alpha).
\end{equation}
For relativistic neutrinos,
\begin{equation}
	P_\nu=\frac{\rho_\nu}{3},
	\qquad
	\rho_\nu+P_\nu=\frac{4}{3}\rho_\nu.
\end{equation}
Thus the neutrino inertial weight entering the momentum-transfer force is enhanced by the familiar factor $4/3$ relative to a pressureless fluid at fixed energy density.

The derivation above separates the covariant consequence of the fluid dynamics, i.e. the transverse pure-momentum-transfer structure and its enthalpy weighting, from the EFT transport closure used to obtain the phenomenological cosmological rate.


\section*{References}
	
	\begin{thebibliography}{99}
\bibitem{Zwicky1933} 
F.~Zwicky, Helv.\ Phys.\ Acta \textbf{6}, 110 (1933).

\bibitem{Riess1998} 
A.~G.~Riess, A.~V.~Filippenko, P.~Challis, A.~Clocchiatti, A.~Diercks, P.~Garnavich, R.~L.~Gilliland, C.~J.~Hogan, S.~Jha, R.~P.~Kirshner \textit{et al.} (High-Z Supernova Search Team), Astron.\ J.\ \textbf{116}, 1009 (1998).

\bibitem{Perlmutter1999} 
S.~Perlmutter, G.~Aldering, G.~Goldhaber, R.~A.~Knop, P.~Nugent, P.~G.~Castro, S.~Deustua, S.~Fabbro, A.~Goobar, D.~E.~Groom \textit{et al.} (Supernova Cosmology Project), Astrophys.\ J.\ \textbf{517}, 565 (1999).

\bibitem{Bertone2005}
G.~Bertone, D.~Hooper, and J.~Silk, Phys.\ Rept.\ \textbf{405}, 279 (2005).

\bibitem{Copeland2006}
E.~J.~Copeland, M.~Sami, and S.~Tsujikawa, Int.\ J.\ Mod.\ Phys.\ D \textbf{15}, 1753 (2006).

\bibitem{Workman2022}
R.~L.~Workman, V.~D.~Burkert, V.~Crede, E.~Klempt, U.~Thoma, L.~Tiator, K.~Agashe, G.~Aielli, B.~C.~Allanach, M Antonelli \textit{et al.}\ (Particle Data Group), Prog.\ Theor.\ Exp.\ Phys.\ \textbf{2022}, 083C01 (2022).
\bibitem{Wetterich1995} C.~Wetterich, Astron.\ Astrophys.\ \textbf{301}, 321 (1995).
\bibitem{Amendola2000} L.~Amendola, Phys.\ Rev.\ D \textbf{62}, 043511 (2000).
\bibitem{Farhan2025} 
M.~Farhan, I.~Hassan, M.~Usman, and N.~Tahir, arXiv:2512.09404.
\bibitem{Simpson2010} F.~Simpson, Phys.\ Rev.\ D \textbf{82}, 083505 (2010).
\bibitem{Farhan:2026sym} M.~Farhan, Phys.\ Scr.\ \textbf{101}, 205302 (2026).
\bibitem{Pourtsidou2013} A.~Pourtsidou, C.~Skordis, and E.~J. Copeland, Phys.\ Rev.\ D \textbf{88}, 083505 (2013).
\bibitem{Coleman1973}
S.~R.~Coleman and E.~Weinberg, Phys.\ Rev.\ D \textbf{7}, 1888 (1973).
\bibitem{Farhan:2026no-go}
M.~Farhan, Eur.\ Phys.\ J.\ C \textbf{86}, 1350 (2026).
\bibitem{Bagherian2026}
H.~Bagherian, M.~Reece, L.-T.~Wang, and H.~Xiao, arXiv:2610.27059.
\bibitem{Koivisto2015} T.~S. Koivisto, E.~N. Saridakis, and N. Tamanini, JCAP \textbf{09}, 047 (2015).
\bibitem{Skordis2015} C.~Skordis, A.~Pourtsidou, and E.~J. Copeland, Phys.\ Rev.\ D \textbf{91}, 083537 (2015).
\bibitem{Pourtsidou2016} A.~Pourtsidou and T.~Tram, Phys.\ Rev.\ D \textbf{94}, 043518 (2016).
\bibitem{Linton2018} M.~S. Linton, A.~Pourtsidou, R.~Crittenden, and R.~Martens, JCAP \textbf{04}, 043 (2018).
\bibitem{Carrilho2022} P. Carrilho, K. Carrion, B. Bose, A. Pourtsidou, J.~C. Hidalgo, L. Lombriser, and M. Baldi, Mon.\ Not.\ Roy.\ Astron.\ Soc.\ \textbf{512}, 3691 (2022).
\bibitem{Bond1980}
J.~R.~Bond, G.~Efstathiou, and J.~Silk, Phys.\ Rev.\ Lett.\ \textbf{45}, 1980 (1980).
\bibitem{Lesgourgues2006}
J.~Lesgourgues and S.~Pastor, Phys.\ Rept.\ \textbf{429}, 307 (2006).
\bibitem{King2015}
S.~F.~King, J.\ Phys.\ G: Nucl.\ Part.\ Phys.\ \textbf{42}, 123001 (2015).
\bibitem{DiracCancellation2009}
S.~F.~King and C.~Luhn, Phys.\ Rev.\ D \textbf{79}, 076007 (2009).
\bibitem{Cao2011}
Q.-H.~Cao, S.~Khalil, E.~Ma, and H.~Okada, Phys.\ Rev.\ Lett.\ \textbf{106}, 131801 (2011).
\bibitem{Bhattacharya2018}
S.~Bhattacharya, I.~de Medeiros Varzielas, B.~Karmakar, S.~F.~King, and A.~Sil, JHEP \textbf{12}, 007 (2018).
\bibitem{Chulia2017}
S.~Centelles Chuli\'a, E.~Ma, R.~Srivastava, and J.~W.~F.~Valle, Phys.\ Lett.\ B \textbf{767}, 209 (2017).
\bibitem{Lattanzi2014}
M.~Lattanzi, R.~A.~Lineros, and M.~Taoso, New J.\ Phys.\ \textbf{16}, 125012 (2014).
\bibitem{Euclid2018}
L.~Amendola, S.~Appleby, D.~Bacon, T.~Baker, M.~Baldi, N.~Bartolo, A.~Blanchard, C.~Bonvin, S.~Borgani, E.~Branchini  et al.\ (Euclid Collaboration)
Living Rev.\ Rel.\ \textbf{16}, 6 (2013).
\bibitem{Planck2018} N. Aghanim, Y. Akrami, M. Ashdown, J. Aumont, C. Baccigalupi, M. Ballardini, A. J. Banday, R. B. Barreiro, N. Bartolo, S. Basak et al. (Planck Collaboration), Astron.\ Astrophys.\ \textbf{641}, A6 (2020).
\bibitem{KiDS2021} C.~Heymans, T.~Tröster, M.~Asgari, C.~Blake, H.~Hildebrandt, B.~Joachimi, K.~Kuijken, C.~Lin, A. G.~Sánchez, J.~Busch et al. (KiDS Collaboration), Astron.\ Astrophys.\ \textbf{646}, A140 (2021).
\bibitem{DESY3} T. M. C. Abbott, M. Aguena, A. Alarcon, S. Allam, O. Alves, A. Amon, F. Andrade-Oliveira, J. Annis, S. Avila, D. Bacon et al. (DES Collaboration), Phys.\ Rev.\ D \textbf{105}, 023520 (2022).
\bibitem{Riess2022} 
A G. Riess, W. Yuan, Lucas M. Macri, D. Scolnic, D. Brout, S. Casertano, D. O. Jones, Y. Murakami, L. Breuval, T G. Brink et al., Astrophys.\ J.\ Lett.\ \textbf{934}, L7 (2022).
\bibitem{Naeem:2026layered} M.~Naeem and M.~Farhan, Phys.\ Scr.\ \textbf{101}, 125301 (2026).
\bibitem{Zu2026}
L.~Zu, W.~Giar\`e, C.~Zhang, E.~Di Valentino, Y.-L.~S.~Tsai, and S.~Trojanowski,
Nature Astron.\ \textbf{10}, 457 (2026).		
\bibitem{PlanckStringBounds}
P. A. R. Ade, N. Aghanim, C. Armitage-Caplan, M. Arnaud, M. Ashdown, F. Atrio-Barandela, J. Aumont, C. Baccigalupi, A. J. Banday, R. B. Barreiro et al.\ (Planck Collaboration), Astron.\ Astrophys.\ \textbf{571}, A25 (2014).
\bibitem{GlobalStringBounds}
M.~Kawasaki, K.~Miyamoto, and K.~Kohri, Phys.\ Rev.\ D \textbf{82}, 103504 (2010).	
\bibitem{Marsh2016}
D.~J.~E.~Marsh, Phys.\ Rept.\ \textbf{643}, 1 (2016).	
\bibitem{SARAH}
F.~Staub, Comput.\ Phys.\ Commun.\ \textbf{185}, 1773 (2014).
\bibitem{deBoer:2020Absolute}
T.~de Boer, M.~Klasen, C.~Rodenbeck, and S.~Zeinstra,
Phys.\ Rev.\ D \textbf{102}, 051702 (2020).
\bibitem{Ma1995}
C.-P.~Ma and E.~Bertschinger, Astrophys.\ J.\ \textbf{455}, 7 (1995).		
\bibitem{Lesgourgues2011} J.~Lesgourgues, arXiv:1104.2932.		
\bibitem{Blas2011} D.~Blas, J.~Lesgourgues, and T.~Tram, JCAP \textbf{07}, 034 (2011).		
\bibitem{Alam2017}
S.~Alam, M.~Ata, S.~Bailey, F.~Beutler, D.~Bizyaev, J.~A.~Blazek, A.~S.~Bolton, J.~R.~Brownstein, A.~Burden, C.~Chuang et al.\ (BOSS Collaboration),
Mon.\ Not.\ Roy.\ Astron.\ Soc.\ \textbf{470}, 2617 (2017).
\bibitem{Bautista2021}
J.~E.~Bautista, R.~Paviot, M.~V.~Magaña, S.~Torre, S.~Fromenteau, H.~Gil-Marín, A.~J.~Ross, E.~Burtin, K.~S.~Dawson, J.~Hou et al.\ (eBOSS Collaboration),
Mon.\ Not.\ Roy.\ Astron.\ Soc.\ \textbf{500}, 736 (2021).
\bibitem{deMattia2021}
A.~de~Mattia, V.~Ruhlmann-Kleider, A.~Raichoor, A.~J.~Ross, A.~Tamone, C.~Zhao, S.~Alam, S.~Avila, E.~Burtin, J.~Bautista et al.\ (eBOSS Collaboration),
Mon.\ Not.\ Roy.\ Astron.\ Soc.\ \textbf{501}, 5616 (2021).
\bibitem{Hou2021}
J. Hou, A. G. Sánchez, A. J. Ross, A. Smith, R. Neveux, J. Bautista, E. Burtin, C. Zhao, R. Scoccimarro, K. S. Dawson et al.\ (eBOSS Collaboration),
Mon.\ Not.\ Roy.\ Astron.\ Soc.\ \textbf{500}, 1201 (2021).
\bibitem{Zhao2019}
G.-B.~Zhao, Y.~Wang, A.~Taruya, W.~Zhang, H.~Gil-Marin, A.~de~Mattia, A.~J.~Ross, A.~Raichoor, C.~Zhao, W.~J.~Percival et al.\ (eBOSS Collaboration),
Mon.\ Not.\ Roy.\ Astron.\ Soc.\ \textbf{482}, 1407 (2019).
\bibitem{Howlett2015}
C.~Howlett, A.~J.~Ross, L.~Samushia, W.~J.~Percival, and A.~Manera, Mon.\ Not.\ Roy.\ Astron.\ Soc.\ \textbf{449}, 848 (2015).
\bibitem{Linton2022} M.~S.~Linton, R.~Crittenden, and A.~Pourtsidou, JCAP \textbf{08}, 075 (2022).
\bibitem{Ma2006}
E.~Ma, Phys.\ Rev.\ D \textbf{73}, 077301 (2006).

			
\end{thebibliography}
\end{document}